\documentclass{emulateapj}
\usepackage{amssymb}
\usepackage{lscape}
\usepackage{longtable}

\begin{document}
\newcommand{\lya}{Lyman~$\alpha$}
\newcommand{\lyb}{Lyman~$\beta$}
\newcommand{\degpoint}{\mbox{$^\circ\mskip-7.0mu.\,$}}
\newcommand{\minpoint}{\mbox{$'\mskip-4.7mu.\mskip0.8mu$}}
\newcommand{\secpoint}{\mbox{$''\mskip-7.6mu.\,$}}
\newcommand{\sqdeg}{\mbox{${\rm deg}^2$}}
\newcommand{\squig}{\sim\!\!}
\newcommand{\subsun}{\mbox{$_{\twelvesy\odot}$}}
\newcommand{\et}{{\it et al.}~}
\newcommand{\Rs}{{\cal R}}

\def\ltsima{$\; \buildrel < \over \sim \;$}
\def\simlt{\lower.5ex\hbox{\ltsima}}
\def\gtsima{$\; \buildrel > \over \sim \;$}
\def\simgt{\lower.5ex\hbox{\gtsima}}
\def\propsima{$\; \buildrel \propto \over \sim \;$}
\def\simprop{\lower.5ex\hbox{\propsima}}
\def\arcs{$''~$}
\def\arcm{$'~$}

\title{THE SPATIAL CLUSTERING OF STAR-FORMING GALAXIES AT REDSHIFTS $1.4\simlt z\simlt 3.5$}

\author{\sc Kurt L. Adelberger\altaffilmark{1}}
\affil{Carnegie Observatories, 813 Santa Barbara St., Pasadena, CA, 91101}

\author{\sc Charles C. Steidel}
\affil{Palomar Observatory, Caltech 105--24, Pasadena, CA 91125}

\author{\sc Max Pettini}
\affil{Institute of Astronomy, Madingley Road, Cambridge CB3 0HA, UK}
                                                                                
\author{\sc Alice E. Shapley\altaffilmark{2}}
\affil{University of California, Department of Astronomy, 601 Campbell Hall, Berkeley, CA 94720}

\author{\sc Naveen A. Reddy \& Dawn K. Erb } 
\affil{Palomar Observatory, Caltech 105--24, Pasadena, CA 91125}

\submitted{Received 2004 June 18; Accepted 2004 October 6}
\shorttitle{CLUSTERING OF STAR-FORMING GALAXIES AT $1.4<Z<3.5$}
\shortauthors{K.L. Adelberger et al.}

\altaffiltext{1}{Carnegie Fellow}
\altaffiltext{2}{Miller Fellow}

\begin{abstract}
We analyzed the spatial distribution of $28500$ photometrically selected
galaxies with magnitude $23.5<{\cal R}_{\rm AB}<25.5$
and  redshift $1.4<z<3.5$ in 21 fields with a total area
of $0.81$ square degrees.  The galaxies were divided into
three subsamples, with mean redshifts
$\bar z=1.7$, $2.2$, $2.9$, according to the $U_nG{\cal R}$
selection criteria of Adelberger et al. (2004) and Steidel et al. (2003).
Combining the galaxies' measured angular clustering with redshift
distributions inferred from $1600$ spectroscopic redshifts,
we find comoving correlation lengths at the three redshifts
of $r_0 = 4.5\pm 0.6$, $4.2\pm 0.5$, and $4.0\pm 0.6 h^{-1}$ Mpc, respectively,
and infer a roughly constant correlation function slope of $\gamma=1.6\pm 0.1$.
We derive similar numbers from the $1600$ object spectroscopic
sample itself with a new statistic, $K$, that is insensitive to many possible
systematics.  Galaxies that are bright in ${\cal R}$ 
($\lambda_{\rm rest}\sim 1500$--$2500$\AA) cluster more strongly than
fainter galaxies at $z=2.9$ and $z=2.2$ but not, apparently,
at $z=1.7$.  Comparison to a numerical simulation that is consistent
with recent WMAP observations suggests that
galaxies in our samples are associated with dark matter halos
of mass 
$10^{11.2}$--$10^{11.8} M_\odot$ ($z=2.9$),
$10^{11.8}$--$10^{12.2} M_\odot$ ($z=2.2$),
$10^{11.9}$--$10^{12.3} M_\odot$ ($z=1.7$),
and that a small fraction of the halos contain more than one galaxy that
satisfies our selection criteria.
Adding recent observations of galaxy clustering at $z\sim 0$ and $z\sim 1$
to the simulation results, we conclude that
the typical object in our samples will evolve into
an elliptical galaxy by redshift $z=0$ and will already have
an early-type spectrum by redshift $z=1$.
We comment briefly on the implied relationship between galaxies
in our survey and those selected with other techniques.
\end{abstract}

\section{INTRODUCTION}
\label{sec:intro}
Early investigators studied the spatial distribution of galaxies
because they hoped to learn about the structure of the universe
on the largest scales.  Their influential work was superseded,
in the end, by its competition.  
Problems began with the demonstration that
galaxies contained only a small fraction of the matter in the universe.
Galaxy formation remained too poorly understood
to quell doubts about  how faithfully galaxies traced
underlying distribution of dark matter.  
Other observations improved---gravitational lensing, peculiar velocities,
intergalactic absorption lines, and so on---and seemed easier to relate
to matter fluctuations.
Computers became fast enough 
to predict the evolution of the large-scale
matter distribution 
from the 
the initial conditions 
that microwave-background missions were measuring with
increasing precision.  
As it became clear that the simulations and observations agreed remarkably
well, most researchers concluded
that the large-scale structure of the universe
could be understood completely as the product of
gravitational instability amplifying small inflationary 
perturbations. 
Galaxies, once believed to be the primary constituent of the universe,
came to be seen as small
test particles 
swept
into ever larger structures
by converging dark matter flows.

The spatial distribution of galaxies remains interesting because
it can teach us about galaxy formation.
Galaxy formation must be closely related to larger process
of gravitational structure formation, 
since the formation of a galaxy begins with gas streaming into
a massive potential well
and ends with stars drifting in the cosmic flow.
Lessons from 20 years of numerical investigations into structure formation
should therefore carry over to the analysis of
galaxy clustering.
One example is the known positive correlation between
clustering strength and mass for virialized dark matter halos.
Since galaxies reside in dark matter halos, their clustering
strength provides an indication of the mass of the halos
that contain them.  The resulting mass estimate depends on the
assumptions that the microwave background and other observations
have given us reliable estimates of cosmological parameters
and of the initial matter power-spectrum,
that numerical simulations can correctly trace the evolution of
the matter distribution, at least for moderate densities,
and that 
no process can significantly separate baryons from dark
matter on Mpc scales---assumptions that are at least as plausible as
those behind competing techniques for mass estimation.
Another example is the evolution in the clustering strength of
a population of galaxies once it has formed.  This is driven
solely by gravity and is easy to predict from numerical
simulations.  Comparing
the clustering of (say) galaxies in the local universe and those
at high redshift can therefore suggest or rule out possible links
between them.  Few other methods are as useful for unifying
into evolutionary sequences
the galaxy populations we observe at different look-back times.
An excellent review of the history of these techniques
has been written by Giavalisco (2002; pp 620-624). 

This paper has two aims.  The first is to present
measurements of the
clustering of UV-selected star-forming galaxies
in a redshift range $1.4<z<3.5$ that is only partially
explored.  Section~\ref{sec:data} describes
the way we obtained our data, \S~\ref{sec:methods}
describes and justifies the techniques we used to estimate the
spatial clustering in our galaxy samples,
and \S~\ref{sec:results} presents our estimates
of the galaxy correlation function at
redshifts $z\sim 1.7$, $z\sim 2.2$, and $z\sim 2.9$.
The survey analyzed here is several times larger
than its predecessors; the surveyed area
of $0.81$ square degrees is roughly 700 times larger
than the area analyzed by Arnouts et al. (2002)
and 4 times larger than the areas analyzed by
Giavalisco \& Dickinson (2001) and Ouchi et al. (2001).
The second aim is to discuss what our measurements
imply about the galaxies and their descendants.
In~\S~\ref{sec:correspondence} we show that
the galaxies' correlation functions are indistinguishable
from those of virialized dark matter halos with mass
$M\sim 10^{12} M_\odot$.  In~\S~\ref{sec:evolution}
we show that the galaxies, dragged by gravity for billions
of years, caught in the press of structure formation, would by redshift
$z=0$ have a correlation function that is indistinguishable
from that of the elliptical galaxies that surround us.
Our results are summarized in~\S~\ref{sec:summary}.

\section{DATA}
\label{sec:data}

\subsection{Observed}
\label{sec:observeddata}
\begin{figure}
\plotone{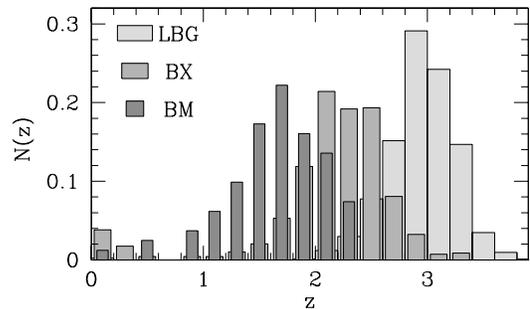}
\caption{
Redshift distributions of objects whose colors satisfy
the three sets of selection criteria presented in \S~\ref{sec:observeddata}.
\label{fig:zhistos}
}
\end{figure}

\begin{figure}
\plotone{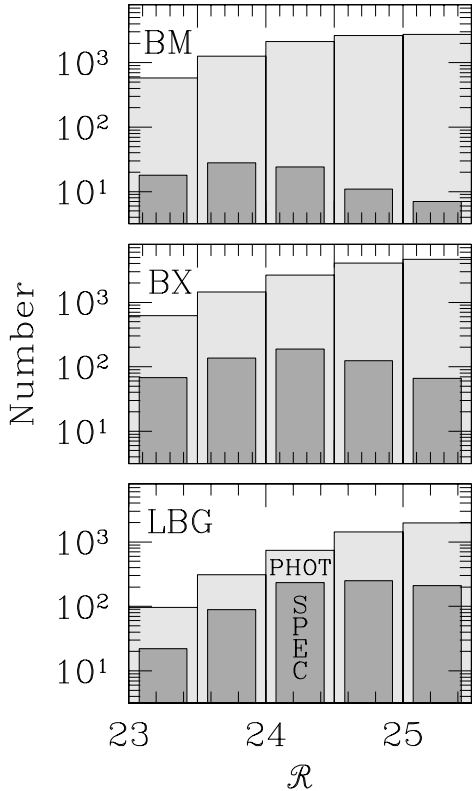}
\caption{
The number of objects in the spectroscopic and photometric samples
as a function of ${\cal R}$ magnitude.
\label{fig:specfrac}
}
\end{figure}

The data we analyzed were drawn from our ongoing surveys of high-redshift
star-forming galaxies.  A brief description of the surveys
follows; see Steidel et al. (2003, 2004) for further details.
Deep, multi-hour $U_nG{\cal R}$ images of 21 fields scattered around the sky
were obtained with various 4m-class telescopes (table~\ref{tab:fields}).  
Tens of thousands of objects were visible in these images.  For the
analysis of this paper we 
ignored all but 
the subset ($\sim$ 20\%) with AB magnitude
\begin{equation}
23.5\leq {\cal R}\leq 25.5
\label{eq:maglimits}
\end{equation}
and AB colors
satisfying the ``LBG'' selection criteria of Steidel et al. (2003),
\begin{eqnarray}
U_n-G     &\geq& G-{\cal R}+1.0\nonumber\\
G-{\cal R}&\leq& 1.2,
\label{eq:lbg}
\end{eqnarray}
the ``BX'' selection criteria of Adelberger et al. (2004),
\begin{eqnarray}
G-{\cal R}&\geq& -0.2\nonumber\\
U_n-G     &\geq& G-{\cal R}+0.2\nonumber\\
G-{\cal R}&\leq& 0.2(U_n-G)+0.4\nonumber\\
U_n-G     &<& G-{\cal R}+1.0,
\label{eq:bx}
\end{eqnarray}
or the ``BM'' selection criteria of Adelberger et al. (2004),
\begin{eqnarray}
G-{\cal R}&\geq& -0.2\nonumber\\
U_n-G     &\geq& G-{\cal R}-0.1\nonumber\\
G-{\cal R}&\leq& 0.2(U_n-G)+0.4\nonumber\\
U_n-G     &<& G-{\cal R}+0.2.
\label{eq:bm}
\end{eqnarray}
In this range of ${\cal R}$ magnitudes, the colors are characteristic
of galaxies at $1.4\simlt z\simlt 3.3$.  (The restriction
to ${\cal R}>23.5$ helps eliminate most interlopers; see
Adelberger et al. 2004 and Steidel et al. 2004.)  By obtaining spectra
of more than $1600$ galaxies with these colors, Steidel et al. (2003, 2004)
established that the mean redshift and $\pm 1\sigma$ range
of galaxies in the three samples is
\begin{equation}
\bar z\pm \sigma_z =\cases{ 2.94\pm 0.30 & LBG \cr
	                      2.24\pm 0.37 & BX  \cr 
	                      1.69\pm 0.36 & BM. \cr}
\label{eq:meanz}
\end{equation}
(These values exclude galaxies with ${\cal R}<23.5$
or ${\cal R}>25.5$ as well as the handful of low-redshift
``interloper'' galaxies with $z<0.8$.)
Redshift histograms are shown in figure~\ref{fig:zhistos}.
We will use the term ``photometric candidates'' to describe
the objects with $23.5\leq {\cal R}\leq 25.5$
whose colors satisfy one set of the selection criteria
presented above, and the term ``spectroscopic sample'' to
describe the subset of photometric candidates that had
a spectroscopic redshift measured by Steidel et al. (2003)
or Steidel et al. (2004).  
Although the spectroscopic
sample is sizeable, it contains only a small fraction ($\simlt 10$\%)
of the photometric candidates (see figure~\ref{fig:specfrac} and
table~\ref{tab:fields}).

\begin{deluxetable*}{lrrrrrrr}
\tablewidth{0pc}
\scriptsize
\tablecaption{Observed fields}
\tablehead{
        \colhead{Name} &
        \colhead{$\Delta\Omega$\tablenotemark{a}} &
        \colhead{$N_{\rm LBG}$\tablenotemark{b}} &
        \colhead{$N_{\rm BX}$\tablenotemark{c}} &
        \colhead{$N_{\rm BM}$\tablenotemark{d}} &
        \colhead{${\cal I}_{\rm LBG}$\tablenotemark{e}} &
        \colhead{${\cal I}_{\rm BX}$\tablenotemark{e}} &
        \colhead{${\cal I}_{\rm BM}$\tablenotemark{e}}
}
\startdata
     3c324& $6.6\times 6.6$  & 11/49   & 0/166    & 0/126  & 0.0035& 0.0034& 0.0042\\
    B20902& $6.3\times 6.5$  & 31/65   & 1/207    & 0/189  & 0.0036& 0.0035& 0.0043\\
      CDFa& $8.7\times 8.9$  & 34/99   & 0/336    & 0/280  & 0.0029& 0.0028& 0.0036\\
      CDFb& $9.0\times 9.1$  & 20/120  & 0/316    & 0/273  & 0.0028& 0.0028& 0.0035\\
  DSF2237a& $9.0\times 9.1$  & 39/100  & 1/367    & 0/328  & 0.0028& 0.0028& 0.0035\\
  DSF2237b& $8.9\times 9.1$  & 44/161  & 1/516    & 0/309  & 0.0028& 0.0028& 0.0035\\
       HDF& $10.4\times 14.4$& 54/187  & 128/735  & 37/587 & 0.0022& 0.0022& 0.0028\\
     Q0201& $8.6\times 8.7$  & 18/90   & 4/339    & 0/285  & 0.0029& 0.0029& 0.0036\\
     Q0256& $8.5\times 8.4$  & 45/126  & 1/346    & 0/243  & 0.0030& 0.0029& 0.0037\\
     Q0302& $15.6\times 15.7$& 46/824  & 0/1778   & 0/749  & 0.0018& 0.0018& 0.0023\\
     Q0933& $8.9\times 9.2$  & 63/192  & 0/435    & 0/273  & 0.0028& 0.0028& 0.0035\\
     Q1307& $16.3\times 16.0$& 16/483  & 47/1352  & 9/936  & 0.0017& 0.0017& 0.0023\\
     Q1422& $7.3\times 15.6$ & 96/253  & 1/728    & 0/491  & 0.0024& 0.0024& 0.0030\\
     Q1623& $12.0\times 22.3$&  6/462  & 189/1220 & 2/847  & 0.0016& 0.0016& 0.0022\\
     Q1700& $15.4\times 15.4$& 15/406  & 62/1456  & 1/948  & 0.0018& 0.0018& 0.0024\\
     Q2233& $9.2\times 9.2$  & 44/76   & 1/267    & 0/181  & 0.0028& 0.0028& 0.0035\\
     Q2343& $22.8\times 11.5$& 10/385  & 148/938  & 8/541  & 0.0016& 0.0016& 0.0022\\
     Q2346& $16.5\times 17.1$&  1/362  & 34/1142  & 1/754  & 0.0016& 0.0017& 0.0022\\
    SSA22a& $8.6\times 8.9$  & 42/151  & 10/360   & 1/253  & 0.0029& 0.0028& 0.0036\\
    SSA22b& $8.6\times 8.9$  & 29/73   & 5/281    & 1/308  & 0.0029& 0.0028& 0.0036\\
  Westphal& $14.9\times 15.1$& 172/270 & 43/724   & 20/632 & 0.0018& 0.0018& 0.0024\\
     Total& 2907             & 836/4934& 676/14009& 80/9533& 0.0021& 0.0021& 0.0029\\
\enddata
\tablenotetext{a}{Area imaged, in square arcmin}
\tablenotetext{b}{Number of sources whose colors satisfy equation~\ref{eq:lbg} in the field's spectroscopic/photometric catalog}
\tablenotetext{c}{Number of sources whose colors satisfy equation~\ref{eq:bx} in the field's spectroscopic/photometric catalog}
\tablenotetext{d}{Number of sources whose colors satisfy equation~\ref{eq:bm} in the field's spectroscopic/photometric catalog}
\tablenotetext{e}{The expectation value of the integral-constraint correction 
for the three samples if galaxies had bias $b=1$.  The 'Total' row
shows the value of equation~\ref{eq:exp_i}.}
\label{tab:fields}
\end{deluxetable*}

All redshifts were measured with the
Low-Resolution Imaging Spectrograph (LRIS; Oke et al. 1995) on the Keck telescopes.
The number of redshifts in each field was determined by the
number of clear nights that were allocated.
Photometric candidates were selected for spectroscopy
more-or-less at random, but in one way
the selection was far from random:
spectroscopic objects in each field were constrained to fit together
in a non-interfering way on one of a small number of multislit masks.
This introduced artificial angular clustering to the spectroscopic
samples, particularly in fields where our image's size significantly
exceeded the spectrograph's $\sim 8'$ field-of-view.
In some cases the artificial angular clustering was increased
by our desire to obtain particularly dense spectroscopic sampling in some
parts of an image, e.g., near a background QSO.

Table~\ref{tab:fields} lists the number of BM, BX,
and LBG photometric candidates and spectroscopic galaxies
in each field.
The field-to-field variations in the number of photometric candidates
per square arcminute
were caused primarily by differences
in exposure times, seeing, sky brightness, telescope plus instrumental
throughput, and so on, from one run to the next.  Recall that
we used many different telescopes and cameras during our imaging survey.
The expected variations in intrinsic surface density (also shown
in the table, and calculated for galaxies with bias $b=1$ 
as described in \S~\ref{sec:angularmethods}
below) are significantly smaller.

We estimated the angular correlation functions from
the lists of photometric candidates.  Inferring a comoving
correlation length $r_0$ from the measured angular clustering
required an estimate of the objects' redshift distribution.
For this we took the measured redshift distributions of
the spectroscopic samples.
Since the spectroscopic samples are large---several hundred
for the BX and LBG criteria, nearly 100 for BM---random fluctuations
are unlikely to have given redshift distributions to them
that are significantly different from those of the parent
photometric samples.  We were able to measure
a redshift for only $\sim 80$\% of the objects we observed
spectroscopically, however, and it is therefore
possible that various systematics (e.g., difficulties
measuring spectroscopic redshifts for galaxies in certain redshift
ranges) could have caused the spectroscopic and photometric
samples to have somewhat different redshift distributions.
Repeated observations of a subset of the initial spectroscopic failures
show that these objects have the same redshift distribution as
the initial successes, implying
that any systematics are not severe.

\subsection{Simulated}
\label{sec:simulated_data}
To help us interpret our observations, we refered at times
the GIF-LCDM numerical simulation
of structure formation in a 
cosmology with
$\Omega_M=0.3$, $\Omega_\Lambda=0.7$,
$h=0.7$, $\Gamma=0.21$, and $\sigma_8=0.9$.
This gravity-only simulation
contained $256^3$ particles with mass $1.4\times 10^{10} h^{-1} M_\odot$
in a periodic cube of comoving side-length $141.3h^{-1}$ Mpc,
used a softening length of $20 h^{-1}$ comoving kpc, and
was released publicly, along with its halo catalogs, by
Frenk et al. astro-ph/0007362.  Further details
can be found in Jenkins et al. (1998) and Kauffmann et al. (1999).
Since the GIF-LCDM cosmology is consistent
with the Wilkinson Microwave-Anisotropy Probe results
(Spergel et al. 2003), and since modeling the gravitational 
growth of perturbations
on large ($\sim$ Mpc) scales is not numerically challenging,
we will assume that the growth of structure found in this simulation
closely mirrors the growth of structure in the actual universe.

For our purposes the most interesting aspect of the simulation
is the spatial distribution of virialized ``halos'', or overdensities
with $\delta\rho/\rho\sim 200$, since these deep potential wells
are the sites where galaxies can form from cooling gas.
The public halo catalogs were created, by the GIF team,
by running a halo-finding algorithm at various time steps
in the simulation.  We will say that all the halos identified
by the algorithm at time-step $t=156$ (say) had $t=156$ as their
time of identification.
Correlation functions for halos at the time of identification
were calculated directly from the public halo catalogs.
In subsequent time-steps these halos were progressively
displaced by gravity. Some grew; others were destroyed as they were
subsumed into larger structures.  Any galaxies within the halos
would be likely to survive intact, however, and it is interesting
to trace the expected evolution in their correlation function
over time.  To do this, we assumed that the galaxies in
a halo would be displaced by gravity by the same amount,
and in the same direction, as the halo's most bound particle.
If $\{p_i\}$ denotes the set of particles that were
the most bound particle in a halo at $t_{\rm earlier}$,
we assumed that the correlation function
at time $t_{\rm later}$ of the galaxies that lay within halos 
identified at time $t_{\rm earlier}$
would be roughly equal to the correlation function of particles $\{p_i\}$
at time $t_{\rm later}$.  
At spatial separations that are large compared to the typical
halo radius,  the expected evolution of the galaxy correlation function
is insensitive to the details of this procedure.  These are
the only spatial separations we will consider.
Our focus on large spatial scales also justifies our
ignoring the possibility of galaxy mergers.  
Although mergers can strongly affect the correlation function on
small scales, on large scales the effect is more subtle.  
It can be understood as follows.  At time
$t_{\rm later}$, the galaxies identified at time $t_{\rm earlier}$
will be found in halos with a range of masses, and
their large-scale correlation function
will be a weighted average of the correlation functions of
the halos.  Since the weighting depends on
typical number of descendant galaxies in halos of each mass,
and since halos with different masses have different correlation
functions,
the large-scale correlation function of descendant galaxies will be altered
by mergers if the merger frequency depends on halo mass.  
In practice, however, the difference in correlation functions
between the more massive and less massive halos that host
the descendants is not enormous, and as a result
the merger frequency would have to be
an implausibly strong function of halo mass to alter
the descendant correlation functions on large scales in a
significant way.

\section{METHODS}
\label{sec:methods}

Two approaches will be used to estimate the clustering strength.
The first approach, which is standard, relies almost exclusively on the
angular positions of the galaxies.  The second relies almost exclusively
on the galaxies' measured redshifts.  These approaches exploit different
aspects of our data and are subject to different systematics.
The level of agreement between them provides an important test of
our conclusions' robustness.  
This section describes and justifies the two approaches.
Readers interested primarily in our
scientific results may wish to skip ahead to \S~\ref{sec:results}.

\subsection{Angular clustering}
\label{sec:angularmethods}

The observed clustering of galaxies on the plane of the sky is
related to the galaxies' three-dimensional correlation function
in a straightforward way.  Let $z$ denote a galaxy's redshift
and ${\bf \Theta}$ denote its angular position.  ${\bf \Theta}$
is written in bold face because two numbers (e.g., right ascension 
and declination) are required to specify the galaxy's angular position.
If $P(z_1{\bf \Theta_1}|z_2{\bf \Theta_2})$ is the probability
that a galaxy at known position $(z_2,{\bf \Theta_2})$ has
a neighbor at position $(z_1,{\bf \Theta_1})$, then
elementary identities show that 
the probability that a galaxy at angular position ${\bf\Theta_2}$
will have a neighbor at angular position ${\bf\Theta_1}$ is
\begin{equation}
P({\bf \Theta_1}|{\bf \Theta_2}) = \int dz_1 dz_2 P(z_2|{\bf \Theta_2}) P(z_1{\bf \Theta_1}|z_2{\bf \Theta_2}).
\label{eq:limber0}
\end{equation}
Observations indicate that the reduced correlation function is 
well approximated by an isotropic power law, 
\begin{equation}
P(z_1{\bf \Theta_1}|z_2{\bf \Theta_2}) \simeq P(z_1{\bf \Theta_1})\bigl[1+(r_{12}/r_0)^{-\gamma}\bigr],
\end{equation}
where $r_0$ and $\gamma$ parametrize the shape of the power law and
$r_{12}$ is the distance between $(z_1,{\bf \Theta_1})$
and $(z_2,{\bf \Theta_2})$. This implies,
in the circumstances of interest to us, that
the reduced angular correlation function will also be a power law,
\begin{equation}
P({\bf \Theta_1}|{\bf \Theta_2})=P({\mathbf \Theta_1})(1+A\theta_{12}^{-\beta})
\end{equation}
with $\beta\equiv\gamma-1$.  
If the angular separation $\theta_{12}\equiv |{\bf\Theta_1}-{\bf\Theta_2}|$
between the
galaxies is small, $\theta_{12}\ll 1$, 
and if the comoving correlation length $r_0$
does not change significantly from the front to the back of the survey,
then 
\begin{eqnarray}
A &=& r_0^\gamma B\biggl(\frac{1}{2},\frac{\gamma-1}{2}\biggr) \int_0^{\infty}dz\,N^2(z)f^{1-\gamma}g^{-1}\nonumber\\
  & &\quad\quad\quad\quad\quad\quad\quad\quad\quad \Bigg/\Biggl[\int_0^{\infty}dz'\,N(z')\Biggr]^2
\label{eq:Alimber}
\end{eqnarray}
(see, e.g., Totsuji \& Kihara 1969)
where $N(z)$ is the survey selection function, $B$ is the beta function
in the convention of Press et al. (1992),
$g(z)\equiv c/H(z)$ is the change in comoving distance with redshift,
$f(z)\equiv (1+z)D_A(z)$ is the change in comoving distance with angle,
and $D_A(z)$ is the angular diameter distance.
This follows from the relationship $\int_{-\infty}^{\infty}dz\, [r_0^2/(R^2+z^2)]^{\gamma/2} = r_0^\gamma R^{1-\gamma} B(1/2,(\gamma-1)/2)$.

Our first approach to estimating the three dimensional clustering strength
will be to measure the parameters
$A$ and $\beta$ of the reduced angular correlation function 
$\omega(\theta)\equiv A\theta^{-\beta}$, 
then infer values for $r_0$ and $\gamma$ using the relationships above.
Our estimates of $\omega(\theta)$
in different angular bins 
will be based on the Landy-Szalay (1993) estimator
\begin{equation}
\omega_{LS}(\theta) \equiv \frac{DD(\theta) - 2DR(\theta) + RR(\theta)}{RR(\theta)}
\label{eq:landyszalay}
\end{equation}
where 
$DD(\theta)$ is the observed number of unique galaxy pairs with separation
$\theta-\Delta\theta/2<\theta<\theta+\Delta\theta/2$, $DR(\theta)$ is the number of unique pairs
with separation in the same range between the observed galaxy catalog
and a galaxy catalog with random angular positions, and $RR(\theta)$
is the number of unique pairs in the random catalog with separations in
the same range.  In practice we reduce the noise in the
random pair counts by creating random catalogs with
many times more objects than the data catalogs 
($n_{\rm rand}/n_{\rm data}\sim 100$),
calculating $DR$ and $RR$, then multiplying $DR$ and $RR$ by
$(n_{\rm data}-1)/n_{\rm rand}$ and $n_{\rm data}(n_{\rm data}-1)/[n_{\rm rand}(n_{\rm rand}-1)]$, respectively.

\subsubsection{Integral constraint}
\label{sec:integralconstraint}
Unless fluctuations on the
size of our typical field-of-view are negligible, the number of
detected galaxies in any field will be somewhat higher or lower than
in a fair sample of the universe, and the number of galaxies
in the field's ideal random catalog would therefore be lower or higher than
the observed number.  As a result the values $DR$ and $RR$ that we calculate
with our approach will be incorrect to some degree.  In a single field
this can make the clustering appear stronger or weaker than it truly is,
but when many fields are averaged it tends to make the observed clustering
appear artificially weak.  This can be shown as follows.
Assume that the observed mean density in a field differs
from the global average by the unknown factor $1+\delta$,
i.e., $\rho_{\rm obs} = \bar\rho (1+\delta)$,
and let $DR$ and $RR$ be the pair counts calculated
from scaling the random catalogs to the observed density.
One might guess that the estimator of equation~\ref{eq:landyszalay}
ought to have $DR$ and $RR$ replaced by values corrected
to the true mean density, i.e., by $DR/(1+\delta)$ and $RR/(1+\delta)^2$,
and indeed
Hamilton (1993, \S 3) has shown that the estimator
\begin{equation}
\omega_{\rm ideal} \equiv \frac{DD - 2DR/(1+\delta) + RR/(1+\delta)^2}{RR/(1+\delta)^2}
\label{eq:hamideal}
\end{equation}
is equal to the true angular correlation function on average,
\begin{equation}
\langle \omega_{\rm ideal}(\theta) \rangle = \omega(\theta).
\end{equation}
This equation does not help us directly, since we do not know $\delta$ and
cannot calculate $\omega_{\rm ideal}$, but it does show that the
estimator $\omega_{LS}(\theta)$ must be biased:
\begin{equation}
\langle \omega_{LS}(\theta) \rangle = \omega(\theta) - \sigma^2\frac{DD}{RR} \simeq \omega(\theta) - \sigma^2
\label{eq:actualestimator}
\end{equation}
where $\sigma^2\equiv{\rm Var}(\delta)$ and 
the approximation assumes the weak clustering ($\omega\ll 1$) limit.
It is therefore customary to estimate $\omega(\theta)$ by adding a constant
${\cal I}\equiv \omega(\theta)-\omega_{LS}(\theta)$
to the calculated values $\omega_{LS}$.  The constant ${\cal I}$
depends on the unknown values of $\delta$ in the observed field or fields
and cannot be calculated exactly.   If $\sigma^2\ll 1$, so that
field-to-field fluctuations are in the linear regime and have a nearly
Gaussian distribution, and if our data are drawn from $n$ independent
fields with measured pair counts $DD_i$, $DR_i$, and $RR_i$, 
then the value of ${\cal I}$ appropriate to a given
angular bin in our data set will have a
variance of 
\begin{eqnarray}
{\rm Var}({\cal I}) &=& \frac{1}{RR_{\rm tot}^2}\sum_{i=1}^n \Bigl[(4\sigma_i^2+2\sigma_i^4)DD_i^2-8\sigma_i^2DD_iDR_i\nonumber\\
		    & & \quad\quad\quad\quad\quad\quad\quad\quad\quad\quad+4\sigma^2DR_i^2\Bigr] \\
               &\simeq& \frac{1}{RR_{\rm tot}^2}\sum_{i=1}^n 2\sigma_i^4 DD_i^2
\label{eq:var_i}
\end{eqnarray}
around its expectation value
\begin{eqnarray}
\langle{\cal I}\rangle &=& \frac{1}{RR_{\rm tot}}\sum_{i=1}^n\sigma_i^2DD_i.
\label{eq:exp_i}
\end{eqnarray}
Here $RR_{\rm tot}\equiv \sum_{i=1}^n RR_i$ is the sum over all fields
of the random pair counts in the chosen angular bin.
In practice $\langle{\cal I}\rangle$ and ${\rm Var}({\cal I})$ depend
very weakly on which bin is chosen.
Below we will take $\langle{\cal I}\rangle$ at $100''$ as our best guess at the 
correction ${\cal I}$.
When it matters we will discuss the effect of the uncertainty in ${\cal I}$.

We use two approaches to estimate the size of
the uncertainty $\sigma_i$
in the mean galaxy density of the $i$th field.
Since
\begin{equation}
\sigma^2 \equiv \frac{1}{\Omega^2}\int_\Omega d\Omega_1\,d\Omega_2\,\omega(\theta_{12}),
\label{eq:integralconstraint}
\end{equation}
it can be estimated numerically as
\begin{equation}
\sigma^2 \simeq \sigma^2_{\rm num}\equiv \frac{\sum_i RR\,\omega(\theta_i)}{\sum_i RR}
\label{eq:rocheeales}
\end{equation}
if $\omega(\theta_i)$ is known (Infante 1994; Roche \& Eales 1999). 
Unfortunately the iterative approach suggested by equations~\ref{eq:actualestimator}
and~\ref{eq:rocheeales} can be unstable, at least when the correlation function
slope is allowed to vary:  a large value of $\omega_{LS}$ will imply a large
correction $\sigma^2$, which implies an even larger $\omega_{LS}$ and even
larger correction, and so on.  The instability is undoubtedly worse
for large images, where the estimate of the integral constraint correction
for one iteration is completely dominated by the assumed correction from the
previous.  We were unable to use
equation~\ref{eq:rocheeales}
as anything other than a consistency check.

A more robust estimate of $\sigma^2$ follows from
theoretical considerations.
Since matter fluctuations will still be in the linear regime on the
large scales of our observations, the relative variance of mass from
one surveyed volume to the next can be estimated from the linear cold-dark matter
power-spectrum $P_L(k)$ (Bardeen et al. 1986; we adopt the
parameters $\Gamma=0.2$, $\sigma_8=0.9$, $n=1$) 
with Parseval's relationship
\begin{equation}
\sigma^2_{\rm CDM} = \frac{1}{(2\pi)^3}\int d^3k P_L(|{\mathbf k}|) |W_k^2({\mathbf k})|
\end{equation}
where $W_k$ is the Fourier transform of a survey volume.  The shape
of the observed volume in any one of our fields is reasonably
approximated in the radial direction by a Gaussian with comoving width (rms) $l_z$ 
and in the transverse directions by a rectangle with
comoving dimensions $l_x\times l_y$.  In this case
\begin{equation}
W_k = \exp\Bigl[-\frac{k_z^2l_z^2}{2}\Bigr] \frac{\sin(k_xl_x/2)}{k_xl_x/2}\frac{\sin(k_yl_y/2)}{k_yl_y/2}.
\end{equation}
The implied value of $\sigma^2_{\rm CDM}$ for each sample in
each field is shown in table~\ref{tab:fields};  
the values assume the powerspectrum normalization required
for the r.m.s. fluctuation as a function of redshift in
spheres of comoving radius $8h^{-1}$ Mpc to
obey $\sigma_8(z)=\sigma_8(0)D_L(z)$ where $\sigma_8(0)=0.9$
and $D_L(z)$ is the linear growth factor to redshift $z$.
The desired
corrections $\sigma^2$ are then given by
\begin{equation}
\sigma^2\simeq b^2 \sigma^2_{\rm CDM}
\label{eq:ic_cdm}
\end{equation}
where $b$, the galaxy bias, is calculated from the ratio of galaxy to matter
fluctuations in spheres of comoving radius $8h^{-1}$ Mpc:
\begin{equation}
b = \frac{\sigma_{8,{\rm g}}}{\sigma_{8}(z)}.
\label{eq:bias_from_sigma}
\end{equation}
Here the galaxy variance
\begin{equation}
\sigma^2_{8,{\rm g}} = \frac{72 (r_0/8 h^{-1}\,\,{\rm Mpc})^\gamma}{(3-\gamma)(4-\gamma)(6-\gamma)2^\gamma}
\label{eq:s2_from_r0}
\end{equation}
(Peebles 1980 eq. 59.3) can be derived from the fit to the galaxy correlation function.
This approach also requires an iterative
solution, since the correction $\sigma^2$ to $\omega(\theta)$
depends on $\omega(\theta)$, but the advantage is that
the assumed size of large-scale fluctuations is anchored in reality by our requirement
that the slope of the correlation function match other observations on
very large scales.  

\subsubsection{Uncertainties in the selection functions}
\label{sec:selfn_uncertainties}
The discussion so far assumes that we will know the precise
shape of the selection function $N(z)$.  In fact this is
not true, and
uncertainty in the true shape of our selection function
is a source of error in the derived values 
of $r_0$.\footnote{
The dependence
of the integral constraint correction on $r_0$ means that
errors in $N(z)$ alter the inferred value of $\gamma$ as well.
We will neglect this small effect.}
A larger width for the selection
function means that projection effects are stronger,
and therefore implies a larger value of $r_0$ for given
angular clustering (see equation~\ref{eq:Alimber}).
If the selection function is a Gaussian with
mean $\mu$ and standard deviation $\sigma_{\rm sel}$, and if the weak
redshift variations of $f$ and $g$ can be ignored, then
the constant $A$
in equation~\ref{eq:Alimber} is proportional
to $\sigma_{\rm sel}^{-1}$,
and the implied value of $r_0$ is proportional to $\sigma_{\rm sel}^{1/\gamma}$.
Measuring $n$ redshifts drawn from this selection function
determines $\mu$ to a precision
$\sigma_{\rm sel}/n^{1/2}$ and $\sigma_{\rm sel}^2$
to a relative precision $2^{1/2}/(n-1)^{1/2}$.
Excluding interlopers with $z<1$,
we have measured roughly 800, 700, and 80 redshifts
for galaxies in the LBG, BX, and BM samples, and 
the selection function width is $\sigma_{\rm sel}\sim 0.3$
for each.  The relative uncertainty in $\sigma_{\rm sel}$ is
therefore approximately $\sim 3$\% for the LBG and BX samples
and $\sim 9$\% for the BM sample, which implies
$\sim 2$\% uncertainty in $r_0$ for the LBG and BX samples
and $\sim 5$\% uncertainty in the BM sample.

Variations of the selection function from one field
to the next (owing, for example, to differences
in the depth of the data or to systematic errors in our
photometric zero points) are another source of concern,
especially at the redshifts $z\sim 2$ where
galaxies' colors are insensitive to redshift
and small color errors mimic large redshift differences.
Suppose for simplicity that
all fields have the same number of photometric candidates,
let the rms width of
the selection function in the $i$th field be
written $\sigma_i=(1+\epsilon_i)s$, where $s$ is the mean width among
all fields, and let the mean redshift of the selection
function be written $\mu+s\delta_i$ where $\mu$ is the mean redshift
among all fields.  Then the rms width of the total selection function
\begin{equation}
[{\rm Var}(z)]^{1/2} = s[1+{\rm Var}(\delta)+{\rm Var}(\epsilon)]^{1/2},
\end{equation}
exceeds the value $s$ that should be used in determining $A$.
Since we will (by necessity) use the total selection function 
in estimating $r_0$,
our estimates will be biased high.  Systematic errors
in our zero points are unlikely to be larger than $\Delta m=0.05$,
and variations in photometric depth will at most change
our characteristic color uncertainties from $\sim 0.1$
to $\sim 0.2$ magnitudes.
The measured variations in galaxy
redshift with $U_nG{\cal R}$ color (see, e.g., Adelberger et al. 2004) 
imply (a) 
that zero point errors with $\Delta m=0.05$ will shift the mean
redshifts of galaxies that satisfy the LBG, BX, and BM selection criteria
by $\Delta z \simeq 0.01$, $0.13$, and $0.11$, respectively,
and (b) that
increasing the photometric uncertainty from $\sigma_m=0.1$ to
$\sigma_m=0.2$ will increase the widths of the LBG, BX, and BM selection 
functions by $\sim 10$, $20$, and $20$\%.
The upper limits on ${\rm Var}(\delta)$ 
and ${\rm Var}(\epsilon)$ are therefore
$0.002$ ($0.2$) and
$0.01$ ($0.04$), respectively, 
for the LBG (BX,BM) sample.
The required reduction in $r_0$ is negligible for the LBG
sample but could be as large as $\sim 7$\% for the other two.
We will account for uncertainties in the selection function
by decreasing the best-guess value of $r_0$ for
the BM and BX samples by 3.5\% and increasing the
uncertainty in quadrature by $0.035r_0$.

\subsubsection{Contaminants}
As figure~\ref{fig:zhistos} shows, some fraction of the objects
in the BX and BM samples will be low redshift interlopers.
We correct for the resulting dilution in the clustering strength
by using the full selection function, starting at $z=0$, in
our estimate of $r_0$ from equation~\ref{eq:Alimber}.
This is the optimal correction only if the interlopers have the
same comoving correlation length $r_0\sim 4h^{-1}$ Mpc
(see below) as the galaxies in the
primary samples.  This should be nearly true, since 
Budav\'ari et al. (2003) estimate $r_0=4.51\pm 0.19 h^{-1}$ Mpc
for the blue star-forming galaxy population at $z\sim 0.2$
from which our interlopers are drawn.  In any case,
since the correction itself is small---eliminating
the tail with $z<1$ from $N(z)$ alters the inferred
values of $r_0$ for the BM and BX samples by only $\sim 10$\%---errors
in it should not have an appreciable effect on
our estimates of $r_0$.

\subsection{Redshift clustering}
\label{sec:redshiftmethod}

We face three significant obstacles in trying to estimate the clustering strength
from the spectroscopic catalogs.  

(1) The objects
in a given field that were selected for spectroscopy 
were not distributed randomly across the field, but were instead
constrained to lie on one of a small number of multislit masks.
Since only a small fraction of the galaxies were observed spectroscopically
in the typical field,
the finite size of the masks coupled with the need to avoid
spectroscopic conflicts produced significant artificial clustering
in the angular positions of sources in the spectroscopic catalog.
The effect was worsened in some fields by our decision to obtain
particularly dense spectroscopy near background QSOs.
(2) Because galaxies' $U_nG{\cal R}$ colors change slowly
with redshift near $z\sim 2$,
the expected redshift distribution $N(z)$ of our 
BM and BX color-selected samples
depends sensitively on the quality of the photometry.
The larger color errors from noisy photometry will lead to
a broader $N(z)$, while relatively small systematic shifts in the
photometric zero points can significantly alter the mean of $N(z)$.
Adelberger et al. (2004) and \S~\ref{sec:selfn_uncertainties} of this
paper discuss this point in more detail, but
the upshot is that we cannot estimate the selection function $N(z)$
with great precision for the BX and BM samples.
(3) Peculiar velocities and redshift uncertainties render imprecise
our estimate of each galaxy's position in the $z$ direction.
This limits the accuracy of our estimate of the distance from one galaxy 
to its neighbors, complicating our efforts to measure the correlation
function on small spatial scales.

Effects (1)--(3) are usually compensated with the aid of detailed
simulations.  Although this approach should work in principle, 
in practice it is hard for outsiders to evaluate whether the
simulations were flawed.  The remainder of this section describes the alternate
approach that we adopt.  It is based on analyzing observable quantities that
are not affected by systematics (1)--(3).

The spurious angular clustering signal can be eliminated if we take
the angular positions of spectroscopic galaxies
as given and estimate the clustering strength solely from
their redshifts.  Let $Z$ be the comoving distance to redshift $z$, and
let ${\mathbf R}\equiv (1+z)D_A(z){\bf\Theta}$ be 
the transverse comoving separation implied by the angular separation 
${\bf\Theta}$ between a galaxy and some reference position, e.g.,
the center of the observed field.
According to elementary probability identities,
if we know that one galaxy has position $({\mathbf R_2},z_2)$,
then the probability that a second galaxy at transverse position
${\mathbf R_1}$ has radial position $Z_1$ is
\begin{equation}
P(Z_1|{\mathbf R_1}{\mathbf R_2}Z_2) = \frac{P(Z_1{\mathbf R_1}|Z_2{\mathbf R_2})}{\int_0^\infty dZ_1' P(Z_1'{\mathbf R_1}|Z_2{\mathbf R_2})}
\label{eq:pzgivenrrz}
\end{equation}
and the expected distribution of radial separations $Z_{12}\equiv Z_1-Z_2$
for galaxies with transverse separation $R_{12}\equiv |{\mathbf R_1}-{\mathbf R_2}|$ is
\begin{eqnarray}
P(Z_{12}|R_{12}) &=& \int_0^{\infty}dZ_2\,P(Z_2|R_{12})P(Z_{12}|Z_2R_{12})\\
                 &\simeq& \bigl[1+\xi(R_{12},Z_{12})\bigr]\times\nonumber\\
 & &\quad \int_0^{\infty} dZ_2\,\frac{P(Z_2)P(Z_2+Z_{12})}{1+P(Z_2)r_0^\gamma R_{12}^{1-\gamma}\beta(\gamma)} 
\label{eq:pzgivenr}
\end{eqnarray}
where we have used results from the previous section and adopted
the shorthand $\beta(\gamma)$ for the beta-function given above, 
$\beta(\gamma)\equiv B(1/2,(\gamma-1)/2)$.  (Equations~\ref{eq:pzgivenrrz}
through~\ref{eq:pzgivenr} assume that the quantity $(1+z)D_A(z)$
is constant with redshift, an approximation that is valid for
the small separations $|Z_{12}|\simlt 40h^{-1}$ comoving Mpc between the galaxy
pairs we will use in this analysis.)

Equation~\ref{eq:pzgivenr} shows that the observable quantity 
$P(Z_{12}|R_{12})$ is sensitive
to the clustering strength but independent of angular variations in the
spectroscopic sampling density.
Unfortunately the correlation function $\xi$ can be estimated from
$P(Z_{12}|R_{12})$ with equation~\ref{eq:pzgivenr} only if we have a reasonably accurate
estimate of the selection function shape $P(Z)$.  
This can be seen more clearly by
Taylor-expanding the integral in equation~\ref{eq:pzgivenr}
around $Z_{12}=0$ and approximating the selection function
as a Gaussian with standard deviation $\sigma_{\rm sel}$
that is centered many standard deviations from $z=0$.
One finds
\begin{eqnarray}
P(Z_{12}|R_{12})&\simeq&\Bigl[1+\xi(R_{12},Z_{12})\Bigr]\times\nonumber\\
                &      &\quad\quad\Bigl[\frac{A_0}{\sigma_{\rm sel}} + \frac{A_2}{2\sigma_{\rm sel}^3}Z_{12}^2+\frac{A_4}{24\sigma_{\rm sel}^5}Z_{12}^4\Bigr]
\label{eq:pzgivenrapprox}
\end{eqnarray}
where 
\begin{equation}
A_n\equiv\int_{-\infty}^{\infty}du\,e^{-u^2}f_n(u)/g(u,R_{12}),\nonumber
\end{equation}
$f_0=1$, $f_2=u^2-1$, $f_4=u^4-6u^2+3$,
$g\equiv 2\pi+(2\pi)^{1/2}a(R_{12})e^{-u^2/2}/\sigma_{\rm sel}$
and $a\equiv r_0^\gamma R_{12}^{1-\gamma}\beta(\gamma)$.
The coefficients $A_n$ all have similar sizes since the integrals
are dominated by contributions from $|u|\simlt 1$ where the
integrands are of the same order.

In the angular clustering case above, inaccuracies in the adopted width 
$\sigma_{\rm sel}$ of the
selection function affected the inferred amplitude of the correlation function
but not its shape.  Here they affect both.
Moreover errors in $\sigma_{\rm sel}$
are multiplied not by $\xi$, but by $1+\xi$, which implies
that they can easily dominate the true clustering signal when $\xi\ll 1$.
Equation~\ref{eq:pzgivenrapprox} shows that one must be careful estimating
the strength of redshift clustering when the shape of the selection function is poorly known.

Our solution exploits the fact that $r_0\ll\sigma_{\rm sel}$ 
for our survey, which implies $Z_{12}\ll\sigma_{\rm sel}$ 
for all separations where $\xi$ is large enough
to measure.  As long as $Z_{12}\ll\sigma_{\rm sel}$, the
terms proportional to $Z_{12}^2$ and $Z_{12}^4$
can be neglected 
in equation~\ref{eq:pzgivenrapprox},
and $P(Z_{12}|R_{12})$ will be very nearly equal to $C(R_{12})[1+\xi(R_{12},Z_{12})]$,
with $C$ a function that does not depend on $Z_{12}$.
The function $C$ does depend on the unknown selection function,
but it can be eliminated by taking ratios of pair counts in a manner
we discuss below.
Ratios of $P(Z_{12}|R_{12})$ at fixed $R_{12}$ and different $Z_{12}$
will therefore be the basis of our estimate of the clustering strength
in the spectroscopic sample; they are nearly immune
to systematics from the irregular spectroscopic sampling and
from the unknown selection function shape.

The final complication is the significant uncertainty $\sigma_Z$ in each
object's radial position $Z$ from peculiar velocities and
redshift uncertainties.  This uncertainty can be treated
in various ways.  We will follow the standard approach 
and estimate the value of the correlation function
only within bins whose radial size $\Delta Z$
is large compared to $\sigma_Z$.

We are now ready to present the estimator that we adopt.
Letting $N(a_1,a_2,R)$ denote the observed number of galaxy pairs
with transverse separation $R$ and
redshift separation $a_1\leq |Z_{12}|<a_2$,
the discussion of the preceding paragraphs shows that
the expected total number of pairs 
with radial separation $a_1\leq |Z_{12}|<a_2$
(and any transverse separation) is
\begin{equation}
\langle N_{\rm tot}(a_1,a_2)\rangle = 2 (a_2-a_1) \sum_{i>j}^{\rm pairs} C(R_{ij}) [1+\bar\xi_{a_1,a_2}(R_{ij})]
\label{eq:expntot}
\end{equation}
where
\begin{equation}
\bar\xi_{a_1,a_2}(R_{ij}) \equiv \frac{1}{a_2-a_1}\int_{a_1}^{a_2} dZ\, \xi(R_{ij},Z).
\end{equation}
As long as $N_{\rm tot}$ is large enough
that 
\begin{equation}
\Biggl\langle\frac{N_{\rm tot}(b_1,b_2)}{N_{\rm tot}(a_1,a_2)+N_{\rm tot}(b_1,b_2)}\Biggr\rangle \simeq \frac{\langle N_{\rm tot}(b_1,b_2)\rangle}{\langle N_{\rm tot}(a_1,a_2)+N_{\rm tot}(b_1,b_2)\rangle},
\end{equation}
the ratio of pair counts
\begin{equation}
K_{a_1,a_2}^{b_1,b_2} \equiv \frac{N_{\rm tot}(b_1,b_2)}{N_{\rm tot}(a_1,a_2)+N_{\rm tot}(b_1,b_2)}
\label{eq:def_k}
\end{equation}
will have expectation value
\begin{eqnarray}
\label{eq:exp_k}
\langle K_{a_1,a_2}^{b_1,b_2}\rangle&\simeq&\frac{N_{\rm exp}(b_1,b_2)}{N_{\rm exp}(a_1,a_2)+N_{\rm exp}(b_1,b_2)}\\
                                    &\simeq&\frac{N'_{\rm exp}(b_1,b_2)}{N'_{\rm exp}(a_1,a_2)+N'_{\rm exp}(b_1,b_2)}
\label{eq:exp_kb}
\end{eqnarray}
regardless of angular selection effects, of uncertainties
in the selection function,\footnote{
More correctly, $\langle K\rangle$ is independent
of uncertainties in the selection function width.
The expectation value $\langle K\rangle$ {\it will}
be affected by errors in the mean redshift of
the selection function if these errors are
large enough to
significantly alter the mapping of redshifts and angles to
distances.}
of peculiar velocities, and of redshift
measurement errors, provided
$a_2\ll\sigma_{\rm sel}$,
$b_2\ll\sigma_{\rm sel}$,
$a_2-a_1\gg \sigma_Z$,
$b_2-b_1\gg \sigma_Z$,
the selection function does not have strong features
on scales smaller than $\sigma_{\rm sel}$,
and $(1+z)D_A(z)$ varies slowly with $z$.
Here $N_{\rm exp}(a_1,a_2)\equiv\langle N_{\rm tot}(a_1,a_2)\rangle$
is given by equation~\ref{eq:expntot} and
\begin{equation}
N'_{\rm exp}(a_1,a_2) = 2 (a_2-a_1) \sum_{i>j}^{\rm pairs} [1+\bar\xi_{a_1,a_2}(R_{ij})].\nonumber
\end{equation}
The second approximate equality in equation~\ref{eq:exp_kb}
exploits the fact that $C(R_{ij})$ is
a very weak function of $R_{ij}$ in realistic situations.
We estimate the correlation function from the spectroscopic
sample by finding the parameters required to match the
observed ratio $K_{a_1,a_2}^{b_1,b_2}$.  In principle
$K_{a_1,a_2}^{b_1,b_2}$ could be calculated separately
for pairs in different bins of transverse separation $R_{ij}$,
producing an estimate of the function $K_{a_1,a_2}^{b_1,b_2}(R)$
and allowing one to estimate both $r_0$ and $\gamma$ from the
data.  In practice a much larger sample is needed to fit
for both $r_0$ and $\gamma$, so we hold $\gamma$ fixed and
estimate $r_0$ only.  Fortunately, as we will see, the best fit
value of $r_0$ hardly changes  as $\gamma$ is varied across the
range allowed by the galaxies' angular clustering.

The dependence of this estimator on the clustering strength is easy to
understand intuitively.  If the galaxies were unclustered ($\xi(r)=0$),
we would observe the same number of pairs at every separation and
$K_{a_1,a_2}^{b_1,b_2}$ would be equal, on average,
to the ratio of the bin sizes $\eta\equiv(b_2-b_1)/(b_2-b_1+a_2-a_1)$.
Correlation functions that peak near $r=0$ will produce more
pairs in bins at smaller separations, driving
$K_{a_1,a_2}^{b_1,b_2}$ away from $\eta$.  The difference between
$K_{a_1,a_2}^{b_1,b_2}$ and $\eta$ is sensitive to the strength of the clustering,
and therefore can be used to estimate it.
Adelberger (2005) uses Monte Carlo simulations to analyze
the behavior of $K_{a_1,a_2}^{b_1,b_2}$ in more detail.

\section{RESULTS}
\label{sec:results}

\subsection{Angular}
\label{sec:angularresults}

\begin{figure}
\plotone{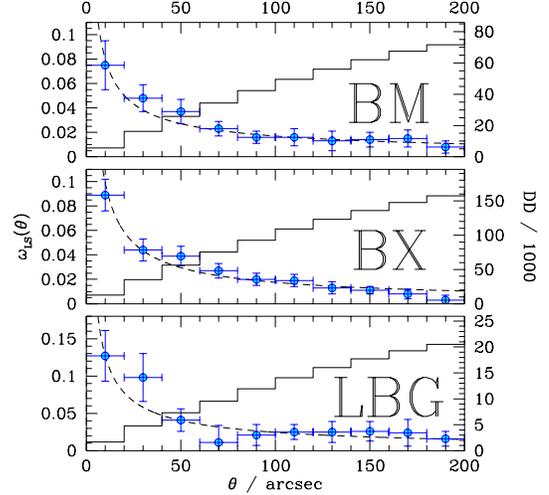}
\caption{
Angular correlation functions uncorrected for the integral constraint
${\cal I}$.  Points with error bars show the measured values
of the estimator $\omega_{LS}$ for each sample.  The dashed line
shows the power-law $\omega=A\theta^{-\beta}$ that fits the
data best.  The solid histogram indicates the number of galaxy
pairs ($/1000$) in each angular bin.
\label{fig:lsraw}
}
\end{figure}

Figure~\ref{fig:lsraw} shows the raw (integral-constraint
correction ${\cal I}=0$) values of the Landy-Szalay estimator
$\omega_{LS}$ (equation~\ref{eq:landyszalay}) as a function of
angular separation for galaxies in the three samples.  
We limited these data, and our subsequent fits, to angular
separations $\theta<200''$, since at larger scales the
weak angular-clustering signal could be swamped by various
low-level systematics.
The uncertainty $\sigma_i$ in each bin was taken to be the larger of
$(DD_i)^{1/2}/(RR_i)$ (Peebles 1980, \S 48)
and the observed standard deviation of the mean
of $\omega_{LS}(\theta_i)$ among the different fields in the survey.
Typically the two were comparable.
Numerical $\chi^2$ minimization produced the
power-law fits
shown with dashed lines.  The correlation function parameters
implied by the LBG fit, $r_0=3.35\pm 0.20 h^{-1}$ comoving Mpc, 
$\gamma=1.74\pm 0.1$, agree well with the estimates of Giavalisco \& Dickinson (2001) which also assumed ${\cal I}=0$.
It is clear, however, that these parameters cannot be correct.  
Substituting
them into equations~\ref{eq:s2_from_r0}, \ref{eq:bias_from_sigma},
and~\ref{eq:ic_cdm} shows that a significant correction ${\cal I}$
should have been applied to account for fluctuations on scales larger
than the field-of-view.  (Porciani \& Giavalisco 2002 reached a similar conclusion,
and derived a result for LBGs that agrees well with the integral-constraint-corrected
result we present below.)

\begin{figure}
\plotone{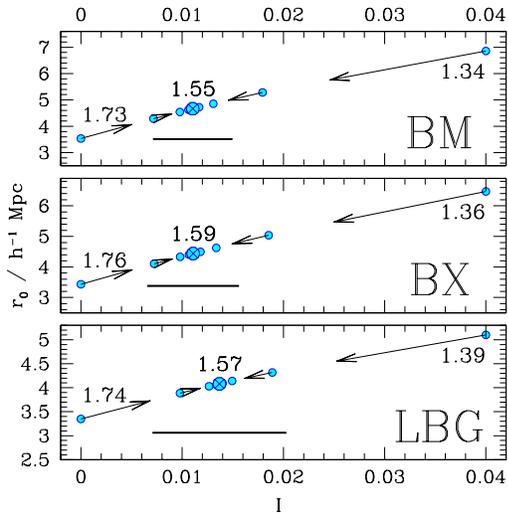}
\caption{
Effect of the integral constraint correction ${\cal I}$
on the inferred correlation function.  The correction depends
on the clustering strength and is therefore unknown initially.
The figure illustrates the iterative approach described in the
text.  Small circles show the adopted value of ${\cal I}$
and the implied value of $r_0$ at each stage in
the iteration.  The numbers above points indicate the value of
$\gamma$ for that iteration; for clarity they are given
for only some iterations.  Adopting the initial guess ${\cal I}=0$
leads to values of $r_0,\gamma$ shown at ${\cal I}=0$ on the plot.
Recalculating ${\cal I}$ for these values of $r_0$,$\gamma$
leads to the revised estimate of the correlation function parameters
shown by the point just above and to the right.  Follow the arrow.   
These new parameters require another update to ${\cal I}$,
which in turn requires a further adjustment to $r_0$ and $\gamma$.
The process monotonically approaches the point of convergence
marked by a large circle enclosing a cross.  The same final
value is reached if the initial guess at ${\cal I}$ is too large.
In general larger corrections ${\cal I}$ imply larger correlation
lengths $r_0$ and shallower slopes $\gamma$.  The $\pm 1\sigma$ uncertainty
in ${\cal I}$ (dark bar in each panel) is comparable to ${\cal I}$
itself, and is a major contributor 
of our total uncertainty.
\label{fig:wthet_converg}
}
\end{figure}

Figure~\ref{fig:wthet_converg} shows how our best-fit estimates of $r_0$ and $\gamma$
change as the correction ${\cal I}$ is applied.  In our first iteration,
described above, we assumed ${\cal I}=0$ and calculated the correlation
function $\omega_1(\theta)$.  For the second iteration we assumed the
value of ${\cal I}$ implied by $\omega_1$ (equations~\ref{eq:exp_i},
\ref{eq:s2_from_r0}, 
\ref{eq:bias_from_sigma}, and~\ref{eq:ic_cdm}) and estimated $\omega_2(\theta)$.
For the third iteration we calculated ${\cal I}$ from $\omega_2(\theta)$.
The process continued in this way until convergence.  It
settled on the same final parameters if we initially
assumed a value for ${\cal I}$ that was too large.
As figure~\ref{fig:wthet_converg} shows, the applied integral
constraint corrections were comparable for each of the three
samples.  This is because the increase in ${\cal I}$ implied
by the longer correlation lengths at lower redshifts
happened to be cancelled by a decrease in ${\cal I}$ that
resulted from the lower-redshift samples' greater comoving depths.

To check the plausibility of our adopted values for ${\cal I}$,
we inserted into equation~\ref{eq:rocheeales} the best power-law fits
to $\omega(\theta)$ from each sample's final iteration.
The equation returned $0.008$, $0.006$, and $0.009$ as the empirical
estimates of ${\cal I}$ for BM, BX, and LBG samples.  These values
differ somewhat from the ones we adopt (figure~\ref{fig:wthet_converg}), 
because the 
empirical and CDM approaches (see~\ref{sec:integralconstraint})
make different assumptions about the behavior of $\omega(\theta)$
on the scales $\theta>200''$ where we cannot measure it,
but they are consistent within their large $1\sigma$ uncertainties
and small changes ($<1\sigma$) to the best-fit values of $\gamma$
would make them agree perfectly.  
Readers may also be reassured to recall that our
estimates of $r_0$ and $\gamma$ agree well
with those of Porciani \& Giavalisco (2002),
who corrected for the integral constraint in a completely
different way.

We estimated the random 
uncertainty in $r_0$ and $\gamma$
in two ways.  First, we analyzed many
alternate realizations of our $\omega(\theta)$ measurements
that were generated under the assumption that the uncertainties
were uncorrelated.
To create a single alternate realization, we added to each
measured value $\omega(\theta_i)$ a Gaussian random
deviate with standard deviation equal to its
uncertainty $\sigma_i$.  After creating numerous alternate realizations,
we calculated and tabulated the values of $r_0$ and $\gamma$
implied by each.  Our $1\sigma$ confidence interval on $r_0$
was defined as the range that contained
68.3\% of the measured values of $r_0$ among the alternate
data sets.
The $\gamma$ confidence interval is defined in the same way.
We found
$r_0= 4.0\pm 0.2 h^{-1}$ Mpc, $\gamma=1.57 \pm 0.07$ (LBG), 
$r_0= 4.3\pm 0.2 h^{-1}$ Mpc, $\gamma=1.59 \pm 0.04$ (BX),
and
$r_0= 4.7\pm 0.2 h^{-1}$ Mpc, $\gamma=1.55 \pm 0.06$ (BM).
These numbers assume uncorrelated error bars and
neglect the uncertainty in our
selection functions.  The uncertainty in ${\cal I}$ is also
neglected, since each alternate realization had the
same integral constraint correction.

Our second approach was to extract random subcatalogs
from our full galaxy catalog, estimate $r_0$ and $\gamma$
for each with the iterative solution for
$\omega(\theta)$ described above, measure how the r.m.s.
dispersion in best-fit parameter values depended
on the number of sources in the subcatalog, 
and extrapolate to the full catalog size.  
In fact we created our random subcatalogs in pairs,
with both subcatalogs in a pair 
containing a random fraction $f\leq 0.5$
of the sources in the full catalog and no sources
in common between them,
and estimated the uncertainty in $r_0$ at a given value of $f$
as $2^{-1/2}$ times the r.m.s. difference in $r_0$ among
pair members.  This prevented us from underestimating
the random uncertainty in $r_0$ as $f\to 0.5$, when
random subcatalogs could otherwise contain nearly
the same galaxies.
With this approach we estimate
$r_0= 4.0\pm 0.5 h^{-1}$ Mpc, $\gamma=1.57 \pm 0.12$ (LBG), 
$r_0= 4.3\pm 0.3 h^{-1}$ Mpc, $\gamma=1.59 \pm 0.05$ (BX),
and
$r_0= 4.7\pm 0.5 h^{-1}$ Mpc, $\gamma=1.55 \pm 0.07$ (BM).
These numbers neglect the uncertainty in our selection function
and do not fully account for the uncertainty in ${\cal I}$.
They do not assume uncorrelated error bars, however, and
we will therefore assume that they are more accurate than the numbers from
the preceding paragraph.

The uncertainty in ${\cal I}$ is not negligible.
According to equation~\ref{eq:var_i} the $1\sigma$ uncertainty
in ${\cal I}$ is $\sim 35$--$50$\% as large as ${\cal I}$ itself
for our 21 fields.  As ${\cal I}$ varies
over its $1\sigma$ allowed range,
the best-fit parameters $r_0$ and $\gamma$ change
by roughly as much as the uncertainties quoted above.
Adding these changes in quadrature to the random uncertainties above,
and making the minor corrections for the selection function
uncertainties discussed in \S~\ref{sec:angularmethods},
we arrive at the following estimates:
$r_0= 4.0\pm 0.6 h^{-1}$ Mpc, $\gamma=1.57 \pm 0.14$ (LBG), 
$r_0= 4.2\pm 0.5 h^{-1}$ Mpc, $\gamma=1.59 \pm 0.08$ (BX),
and
$r_0= 4.5\pm 0.6 h^{-1}$ Mpc, $\gamma=1.55 \pm 0.10$ (BM).

\begin{figure}
\plotone{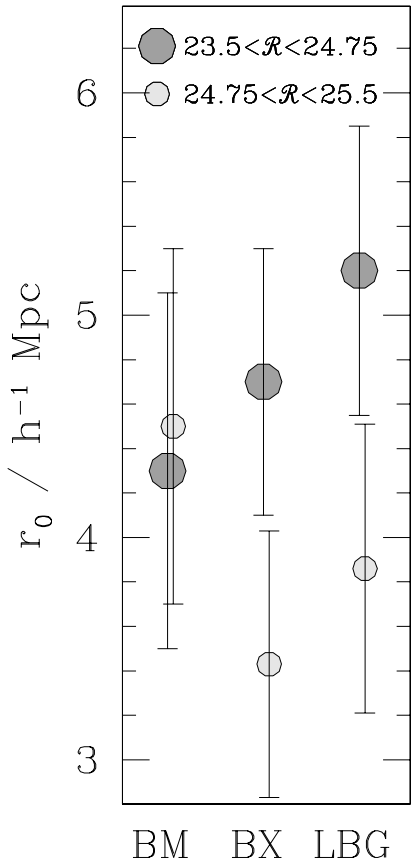}
\caption{
Correlation length $r_0$ for bright and faint subsamples
of the BM, BX, and LBG samples.
\label{fig:clust_seg}
}
\end{figure}

Other investigators (e.g., Giavalisco \& Dickinson 2001;
Foucaud et al. 2003) have claimed that  at redshift $z\sim 3$
bright galaxies cluster more strongly than faint
galaxies.  Our data support this conclusion.
Figure~\ref{fig:clust_seg} shows that in the BX and LBG samples
the correlation lengths of galaxies with
$23.5<{\cal R}<24.75$ ($r_0\sim 5.0h^{-1}$ Mpc)
exceed those of galaxies with $24.75<{\cal R}<25.5$ ($r_0\sim 3.7h^{-1}$ Mpc)
by a significant amount.
If we split the BX and LBG samples into two halves
at random, rather than by apparent magnitude, the difference
in correlation lengths between the two halves is this large
only about $5$\% (BX) to $24$\% (LBG) of the time.
The situation is less clear for the BM sample at $z\sim 1.7$,
where the uncertainties are larger
owing to the poor determination of $N(z)$
from the small number of measured redshifts,
but the data do not seem to suggest stronger clustering
for brighter galaxies.  On the one hand it makes
sense that 
UV-brightness should become less associated with strong clustering
as redshift decreases, since UV-bright galaxies are
known to be weakly clustered at $z=0$ and $z=1$.
On the other, the overall clustering of the BM sample
is still quite strong, stronger than one expects for typical
collapsed objects at $z\sim 1.7$ (see below), and 
so it seems that the numerous objects too faint to satisfy our selection
criteria must be less clustered than the bright objects in our sample.
We will wait for additional spectroscopic observations of BM galaxies
before commenting further.

The dependence of clustering strength on luminosity
can produce a false impression of a change in $r_0$ with
redshift, since lower redshift samples will tend to 
reach fainter absolute luminosities.  Truncating the samples
at a fixed absolute luminosity does not seem a good
solution to us, however, since the bright end of the UV luminosity
function rises rapidly towards higher redshifts (e.g., Adelberger \& Steidel 2000)
and one would therefore be comparing rare objects at lower
redshifts to common objects at higher redshifts.  A better approach
is to compare galaxy samples of roughly the same
comoving number density.  Since selection
with a constant apparent magnitude limit ${\cal R}<25.5$
happens to produce similar comoving number densities for the three samples
(see equations~\ref{eq:nlbg} and~\ref{eq:nbxbm} and the 
related discussion), we will continue to use
the constant apparent magnitude limits of equation~\ref{eq:maglimits}
for our samples in the remainder of the paper. Readers should be aware
that the reported value of each sample's correlation length
is somewhat arbitrary for this reason.  It reflects the characteristics
of the sample as defined here, not of the general galaxy population
at high redshift.

\subsection{Redshift}

For our estimator of the redshift clustering strength
we took $K^{0,20}_{20,40}$, the ratio 
of the number of galaxy pairs with comoving radial separation $0<|Z|/(h^{-1}{\rm Mpc})<20$
to those with comoving radial separation 
$0<|Z|/(h^{-1}{\rm Mpc})<40$.
Since $20 h^{-1}$ Mpc is significantly larger
than the uncertainty in each galaxy's radial position
($\sigma_Z\simeq 300\,{\rm km}{\rm s}^{-1}(1+z)/H(z)\sim 3h^{-1}$ comoving Mpc),
and since $40 h^{-1}$ Mpc is significantly smaller
than the selection functions' widths 
($\sigma_{\rm sel}>200 h^{-1}$ comoving Mpc),
the expected value of $K^{0,20}_{20,40}$
should be given by equation~\ref{eq:exp_k}.
We limited our analysis to pairs with transverse
separations $\theta_{ij}<300''$, equivalent to 
$R_{ij}\simlt 5.9h^{-1}$ comoving Mpc
at $z=2.5$, to reduce the sensitivity of our results
to any deviations of the correlation function
from a $\gamma=1.55$ power-law on large scales.

Only the BX and LBG spectroscopic samples were large
enough to allow meaningful measurements of $K^{0,20}_{20,40}$.
For $\gamma=1.55$, the right-hand side of equation~\ref{eq:exp_k} 
is equal to the observed ratio $K^{0,20}_{20,40}$
when $r_0=4.6h^{-1}$ (LBG) or $4.5h^{-1}$ (BX) comoving Mpc.
The values change by roughly $\pm 2$\% as $\gamma$
is varied from 1.45 to 1.65.
When analyzed with this technique, 
mock galaxy catalogs from the GIF simulation (\S~\ref{sec:simulated_data})
with sizes similar to our observed catalogs show
a $\pm 1\sigma$ dispersion in $r_0$ around the true mean
of $0.6 h^{-1}$ (LBG) and $0.9 h^{-1}$ (BX) comoving Mpc,
so we adopt $4.6\pm 0.6 h^{-1}$ and $4.5\pm 0.9 h^{-1}$ Mpc
as the best fit values to $r_0$ for our spectroscopic catalogs.
The results do not change significantly if we eliminate pairs
with $\theta_{ij}<60''$ from the analysis, showing that we
have measured genuine large-scale clustering and not merely
the clumping of objects within individual halos.

\begin{figure}
\plotone{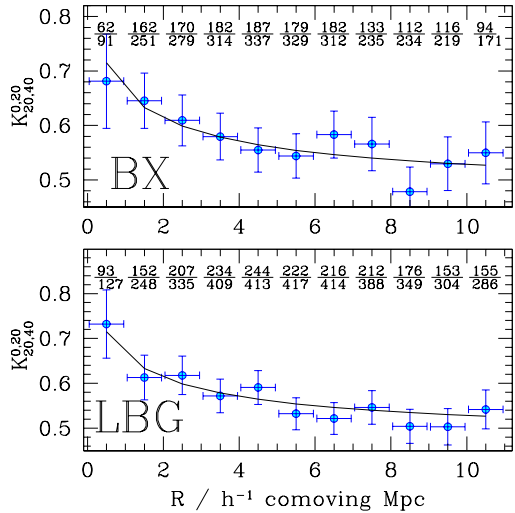}
\caption{
Redshift clustering in the BX and LBG samples.  The ordinate
shows $K_{20,40}^{0,20}$, the fraction of galaxy
pairs with comoving radial separation $0<Z<40 h^{-1}$ Mpc that
have $0<Z<20 h^{-1}$ Mpc.  $K_{20,40}^{0,20}$
was calculated for galaxy pairs in different
bins of transverse separation $R$.
In the absence of
clustering $K_{20,40}^{0,20}$ would be
equal to $0.5$, on average, at all
values of $R$.  Instead our data (points with 
error bars) show that $K_{20,40}^{0,20}$ is
generally larger than $0.5$, especially at smaller $R$.
This implies a positive correlation function that rises  
towards smaller separations.  The fractions at the top
of the plots show the number of galaxies
with $0<Z<20 h^{-1}$ Mpc and $0<Z<40 h^{-1}$ Mpc
in each bin of $R$.  The error bars are equal to
$n^{1/2}/d$ where $n$ is the numerator and $d$ the denominator in the fraction.
They are only approximate; the actual analysis treated uncertainties
as described in the text.
The solid line shows the values of $K_{20,40}^{0,20}$ implied
by the best fit of a power-law correlation function 
$\xi(r)=(r/r_0)^{-1.55}$ to the data when they are placed
in a single radial bin.  
\label{fig:kclust}
}
\end{figure}

Figure~\ref{fig:kclust} presents the result in a more graphical way.
We divided our lists of galaxy pairs into bins according
to transverse separation $R_{ij}$, then calculated $K^{0,20}_{20,40}$
separately for each bin.  Points with error bars show the values we found.
The solid lines show the values predicted by
the $\gamma=1.55$ correlation function described in the preceding
paragraph.  The plot shows that the derived correlation function
parameters provide a reasonable fit to the clustering of
the galaxies in the spectroscopic sample.

\subsection{Summary}
We presented two independent estimates of the correlation function
for each of our galaxy samples.  The estimates were consistent with
each other, but the first, based on the galaxies' angular
clustering, had somewhat smaller uncertainties.  
This resulted from the larger size of the photometric
sample, and was accentuated by the serious systematics
in the spectroscopic sample that made us throw much of our data away.
We will adopt the angular clustering results
for the remainder of the paper.

\section{IMPLICATIONS}

\subsection{Correspondence to halos}
\label{sec:correspondence}
On small scales, smaller than roughly the typical radius 
$r_{\rm vir}$ of a virialized halo, 
the spatial clustering of galaxies is difficult to
predict or interpret.  It depends on the ease with which
nearby galaxies merge with each other, on the ability of
a galaxy to maintain its star-formation rate as it 
orbits within a larger potential well, 
on the possible impact of a galaxy's
feed-back on its surroundings, and so on.
On larger scales these baryonic complications have little effect 
and the correlation function of galaxies 
should be virtually identical to the correlation function
of the halos that host them.
To see that this is true,
consider the galaxy correlation function in a simplified 
situation where
every galaxy is associated with a single halo and the probability
that a galaxy lies a distance $r$ from its halo's center is
$f(r)$.  In this case the galaxy distribution will be a Poisson
realization of the continuous density field $\rho_g({\mathbf r})$
that is created when the discrete halo distribution is convolved by $f$,
and the galaxy correlation function $\xi_g$ will be equal to the
correlation function of $\rho_g$ (see, e.g., Peebles 1980, eq. 33.6).  
Since the halo powerspectrum is
$P_h({\mathbf k}) = 1/n + \int d^3r \xi_h \exp(i{\mathbf k}\cdot{\mathbf r})$
(Peebles 1980, eq. 41.5),
where $\xi_h$ is the halo correlation function and $n$ is
the halo number density,  since the powerspectrum of $\rho_g$ is equal
to $|\tilde f({\mathbf k})|^2P_h({\mathbf k})$, where 
$\tilde f$ is the Fourier transform of $f$, 
and since the powerspectrum and correlation function are
Fourier-transform pairs, 
the galaxy and halo correlation functions will be related through
\begin{equation}
\xi_g({\mathbf r}) = F({\mathbf r})/n + F({\mathbf r})\otimes \xi_h({\mathbf r})
\label{eq:xigxih}
\end{equation}
where $\otimes$ denotes convolution and $F\equiv f\otimes f$.
Now $f(r)=0$ when $r$ is greater than some
maximum separation $r_{\rm max}\sim r_{\rm vir}$; galaxies cannot be located
arbitrarily far from the center of their halo.   
The first term will therefore be zero for $r\gg 2r_{\rm vir}$.
The second term will be
almost exactly equal to $\xi_h(r)$ at the same large
separations, because plausible correlation functions do not
not change significantly from $r$ to $r+2r_{\rm vir}$
when $r_{\rm vir}\ll r$.  
This shows that $\xi_g(r)\simeq \xi_h(r)$ for large $r$.
Although it was derived for a simplistic model, the result
is more general.  As long as galaxies are associated in some way with
dark matter halos, and as long as there is some maximum separation
$r_{\rm max}\sim r_{\rm vir}$ between each galaxy and the center of its halo,
galaxies will have the same correlation function as the halos that
host them for $r\gg r_{\rm vir}$.
We will accordingly focus 
our attention 
solely on separations $r\simgt 1h^{-1}$ comoving Mpc
that are many times larger than the typical virial radius.

\begin{figure}
\plotone{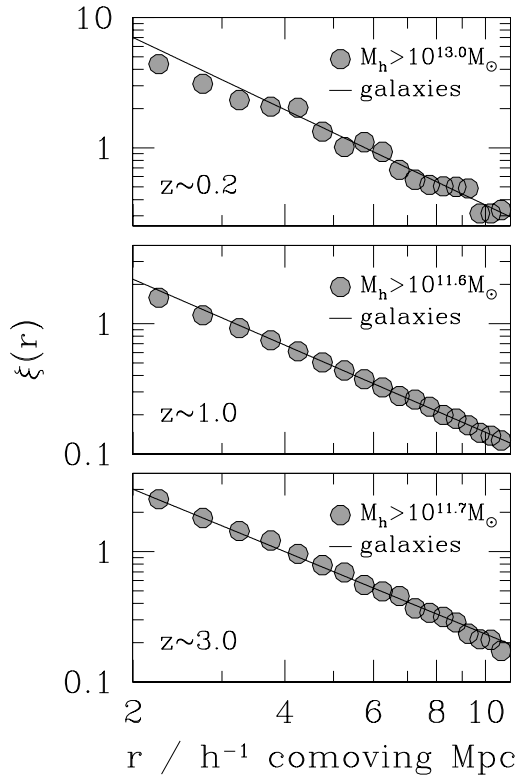}
\caption{
Comparison of observed galaxy correlation functions
to the expected correlation functions of dark matter halos.
Solid lines show
the power-law fits implied by the parameters
reported by Budav\'ari et al. (2003; $r_0=5.77\pm 0.09 $, $\gamma=1.84\pm 0.02$, $z\simeq 0.2$),
by Coil et al. (2003; $r_0=3.19\pm 0.51$, $\gamma=1.68\pm 0.07$, $z\simeq 1.0$),
and by us (\S~\ref{sec:results}, $z\simeq 3.0$).  In each sample the
galaxies' clustering was measured out to a maximum transverse separation
of roughly $10 h^{-1}$ comoving Mpc.  Circles show the correlation
function of halos in the LCDM-GIF simulation
(see \S~\ref{sec:simulated_data}).  The halo correlation functions
were calculated at the redshift indicated in each panel,
and only halos with masses larger than the indicated threshold
were included.  The comparatively small number of halos with $M>10^{13} M_\odot$
is responsible for the noisier halo correlation function at $z=0.2$.
In general the agreement between the galaxy and halo correlation functions
is excellent, although
differences between the galaxy and halo correlation functions
arise as $r\to 2 h^{-1}$ Mpc and $z\to 0$ because $r$
is no longer enormous compared to the halo radius.
\label{fig:xi_halo_0z3}
}
\end{figure}

Readers who are skeptical of the claimed similarity between
galaxy and halo correlation functions at $r\gg r_{\rm vir}$
may wish to consider figure~\ref{fig:xi_halo_0z3},
which compares observed galaxy correlation functions
at redshifts $z=0.2$, $1.0$, $3.0$ to the halo correlation functions
in the GIF-LCDM simulation outputs at the same redshifts
(see~\S~\ref{sec:simulated_data}).  The agreement is excellent.

\begin{figure}
\plotone{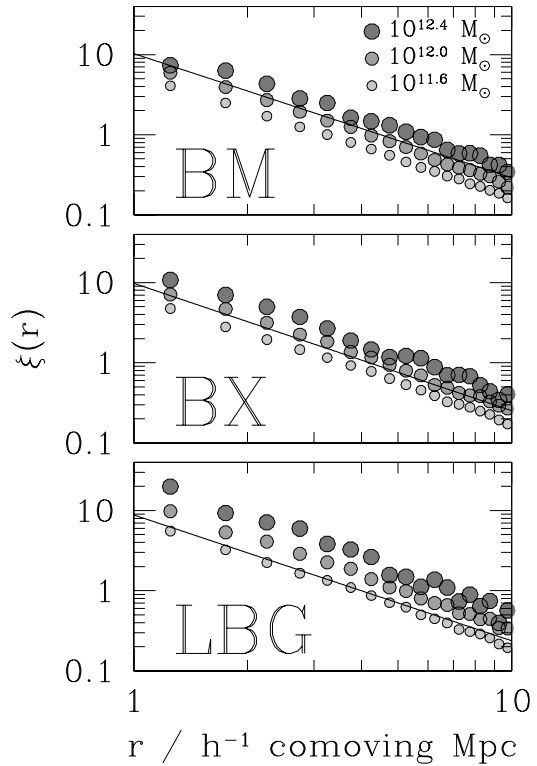}
\caption{
Our best correlation function fits (solid lines) compared to the
correlation functions of virialized dark matter halos 
in the GIF simulation.  The halo correlation functions were
calculated for halos with mass $M>10^{11.4} M_\odot$,
$M>10^{12.0} M_\odot$, and $M>10^{12.6} M_\odot$
that were identified at redshifts
$z=1.77$ (BM), $2.32$ (BX), and~$2.97$ (LBG).
The dependence of halo clustering strength on mass
allows one to estimate characteristic masses of
the halos that are associated with the galaxies in our samples.
\label{fig:xi_halo}
}
\end{figure}

The implied association of galaxies with 
massive potential wells
is hardly surprising.
The interesting result is the characteristic mass of
the virialized halos that contain the galaxies.  This can
be estimated since more massive halos cluster
more strongly.
Figure~\ref{fig:xi_halo} compares the correlation functions
at $1\simlt r/(h^{-1}{\rm comoving\ Mpc}) \simlt 10$
for galaxies in the BM, BX, and LBG samples to the 
correlation functions of virialized dark matter halos above various mass thresholds
in the GIF-LCDM simulation (see~\S~\ref{sec:simulated_data}).  
The agreement is best if galaxies in the BM, BX, and LBG samples
are associated with halos of mass $M\sim 10^{12.1} M_\odot$, 
$10^{12.0} M_\odot$, and $10^{11.5} M_\odot$, respectively.

\begin{figure}
\plotone{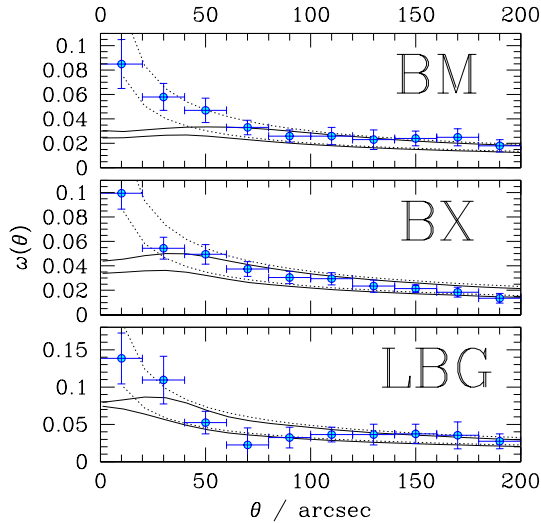}
\caption{
Observed galaxy angular correlation functions compared to
the angular correlation functions of halos in the GIF-LCDM 
simulation.  Points with error bars show our observations,
including integral constraint correction ${\cal I}\sim 0.01$.
(Figure~\ref{fig:wthet_converg} shows the precise
values of ${\cal I}$ that were adopted.)  Solid lines show
the halo angular correlation functions for GIF-LCDM
halos of mass $M=10^{11.9}$, $10^{12.3} M_\odot$
at the identification redshift $z=1.77$ (BM),
of mass $M=10^{11.8}$, $10^{12.2} M_\odot$
at the identification redshift $z=2.32$ (BX),
and of mass $M=10^{11.2}$, $10^{11.8} M_\odot$
at the identification redshift $z=2.97$ (LBG).
The angular correlation functions were calculated for
the redshift distributions shown in figure~\ref{fig:zhistos}.
In each case the curves for the more massive halo are on top.
The plateau in the halo angular clustering at small radii
results from the definition of a halo, which
imposes a minimum separation for halo neighbors by counting
two halos as the same object if they lie too close together.
Dotted lines show the angular correlation functions that
result when the power-law fits to the three-dimensional
halo correlation functions
at $2<r<10h^{-1}$ comoving Mpc are extrapolated to $r=0$.
They are for reference only;
the galaxy and halo angular correlation functions should be compared only
at the separations $\theta\simgt 60''$ where they are
expected to be similar (see equation~\ref{eq:xigxih}).
The two halo curves in each panel roughly span the range
in angular clustering strength allowed by the
uncertainty of $\sim 0.005$ in ${\cal I}$, and
so the masses of the halos listed above should roughly bracket
the allowed masses of the halos that host the galaxies.
\label{fig:wthet_halo}
}
\end{figure}

Uncertainties in our measured correlation functions lead to
uncertainties in the estimated characteristic masses.
The size of these uncertainties can be gauged most cleanly
by comparing our measured angular correlation function
to the angular correlation functions of halos above different
mass thresholds.  We calculated halo angular correlation functions
numerically from equation~\ref{eq:limber0} after substituting in
the observed redshift distributions for
the different samples (fig.~\ref{fig:zhistos}) and
the GIF-LCDM halo correlation functions at the mean redshift
of each sample.  Typical results are shown in figure~\ref{fig:wthet_halo}.
The halo angular correlation functions fall significantly
below the data on small scales, since (by definition)
a halo cannot have another halo as a neighbor within
the radius $r_{\rm vir}$.  As argued above, however,  it
is the larger scales $r\simgt 1h^{-1}$ comoving Mpc
(i.e., $\theta\simgt 1'$) that are relevant for
comparing to galaxies, and here the agreement is good.
The uncertainty in the galaxies'
implied mass scale is dominated by the uncertainty in
the integral constraint correction ${\cal I}$ that was applied to the
data, since this moves all points up or down together.
Rough $1\sigma$ limits on the threshold masses of the hosting
halos are 
\begin{eqnarray}
\label{eq:mass_scales}
10^{11.9}  &\simlt M/M_\odot\simlt& 10^{12.3}\quad\quad{\rm BM}\nonumber\\
10^{11.8}&\simlt M/M_\odot\simlt& 10^{12.2}\quad\quad{\rm BX}\\
10^{11.2}&\simlt M/M_\odot\simlt& 10^{11.8}\quad\quad{\rm LBG}.\nonumber
\end{eqnarray}
Readers unfamiliar with the idea of threshold masses may wish to
see footnote~6 in~\S\ref{sec:summary}, below, for further discussion
of how to interpret them.

One implication of figure~\ref{fig:wthet_halo} is that
some halos are occupied by more than one of the galaxies
in our samples; the data on small scales are strongly inconsistent with
the correlation functions that assumed one object per halo.
(This was first pointed out by Wechsler et al. (2001) and later denied
by Porciani \& Giavalisco (2002).  Our analysis agrees far better with that of
Wechsler et al.)
The fraction of LBGs that reside in the same halo as another
LBG can be calculated as follows.  If there were never more than
one LBG per halo, the expected number of LBGs within $1'$ of a
randomly chosen
LBG would be $N_1={\cal N}(1+\bar\omega_1)$ where
${\cal N}$ is the surface density of LBGs,
$\bar\omega_1\equiv \int_0^{1'}\theta d\theta\,\omega_h(\theta)/\int_0^{1'}\theta d\theta$,
and $\omega_h$ is halo angular correlation function that fits
our data best for $\theta>1'$.
The actual number of LBGs within $1'$, $N_{\rm true}$, is
given by the same equation with the halo correlation function
$\omega_h$ replaced by the galaxy correlation function
$\omega_{\rm true}$. 
The expected number of additional LBGs in a halo that is known
to contain one LBG, $f_{\rm LBG}$, is equal to the
difference between $N_{\rm true}$ and $N_1$.
Numerically integrating the angular correlation functions
for observed galaxies and simulated halos, and multiplying
by the LBG surface density from table~\ref{tab:fields}, we estimate
$f_{\rm LBG}\sim 0.05$.  The numbers for the other samples,
estimated with a similar approach,
are $f_{\rm BX}\sim 0.25$ and $f_{\rm BM}\sim 0.25$.
{\it Some} galaxies must share the same halo to explain the data,
but the required number is small.  Note that we are referring
solely to galaxies that satisfy our color and magnitude selection
criteria; we obviously cannot say anything about the spatial distribution
of galaxies that are not in our samples.

\begin{figure}
\plotone{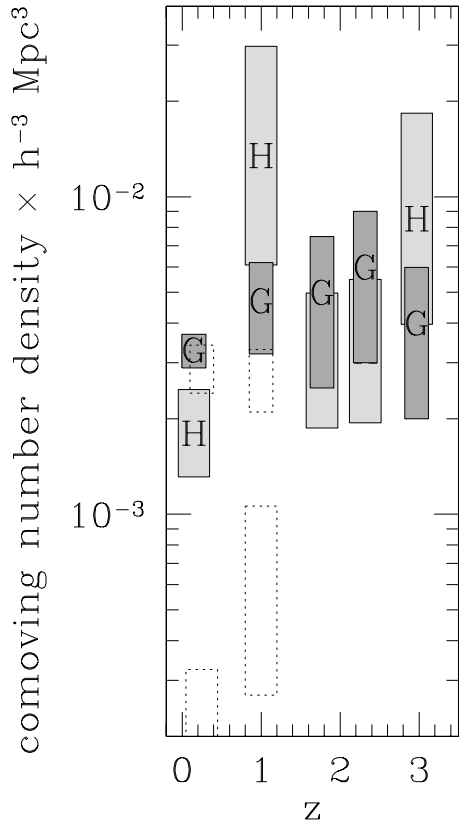}
\caption{
The comoving number densities of galaxies (G) in various samples
compared to the comoving number density of the halos (H) 
at the same redshift that have the same clustering
strength (on large scales) as the galaxies.  These are the halos
that the galaxies can reside within.  Shaded
regions are for star-forming galaxies and
their associated halos:  the ``blue'' SDSS sample
of Budav\'ari et al. (2003; $z=0.2$), the emission-line sample
of Coil et al. (2004; $z=1$), and our three samples
($z>1.5$).  The height of the rectangles covers the $\pm1\sigma$ range
in number density.  The uncertainties in halo number density
follow from the uncertainty in their galaxy's clustering strength.
At each redshift star-forming galaxies are roughly as numerous
as the halos that can host them.
Galaxies that are no longer forming stars tend to outnumber
halos with similar clustering strength, which shows that
several reside in the same halo.  This is evident
in both the absorption-line sample of Coil et al. (2004;
dotted rectangles at $z=1$) and 
the elliptical sample of Budav\'ari et al. (2003; $-23<M_r<-21$,
dotted rectangles at $z=0.2$).  For consistency with
the Budav\'ari sample, the number density of early-type galaxies
at $z\sim 1$ was calculated by integrating the $z\sim 1$
early-type luminosity function
of Chen et al. (2003) over all magnitudes brighter than $M_R=-21$.
\label{fig:nhalo}
}
\end{figure}

Having established a rough characteristic mass for the halos
that contain our galaxies, we can compare the galaxy and halo number
densities and estimate what fraction of the most massive halos
at redshifts $1.5\simlt z\simlt 3.5$ do not contain a galaxy
that is detectable with our techniques.  According to
Adelberger \& Steidel (2000)
the comoving number density of LBGs
brighter than ${\cal R}=25.5$ is 
\begin{equation}
n_{\rm LBG}=(4\pm 2)\times 10^{-3} h^{3} {\rm Mpc}^{-3}
\label{eq:nlbg}
\end{equation}
for $\Omega_M=0.3$, $\Omega_\Lambda=0.7$.
Combining the BM and BX completeness coefficients in table~3
of Adelberger et al. (2004) with the surface densities in
table~\ref{tab:fields} of this paper, we estimate
\begin{eqnarray}
n_{\rm BX}&=&(6\pm 3)\times 10^{-3} h^{3} {\rm Mpc}^{-3}\nonumber\\
n_{\rm BM}&=&(5\pm 2.5)\times 10^{-3} h^{3} {\rm Mpc}^{-3}
\label{eq:nbxbm}
\end{eqnarray}
where the assigned uncertainties of 50\% are approximate guesses
that will be refined later with Monte Carlo simulations.
Number densities for the population of halos that contain the galaxies
can be estimated from the GIF-LCDM simulation given the
range of halo masses (equation~\ref{eq:mass_scales})
that are compatible with the galaxies'
clustering strength.  Figure~\ref{fig:nhalo} shows that
the number density of galaxies in the BM, BX, and LBG samples
is comparable to the number density of halos that can host them.
As we will discuss in \S~\ref{sec:summary}, below,
this implies that the duty cycle of star-formation in the galaxies
must be of order unity and shows that our surveys cannot be severely
incomplete.  Similar arguments have been made by
Adelberger et al. (1998), Giavalisco \& Dickinson (2000),
Martini \& Weinberg (2001), and others.

\subsection{Evolution}
\label{sec:evolution}
\begin{figure}
\plotone{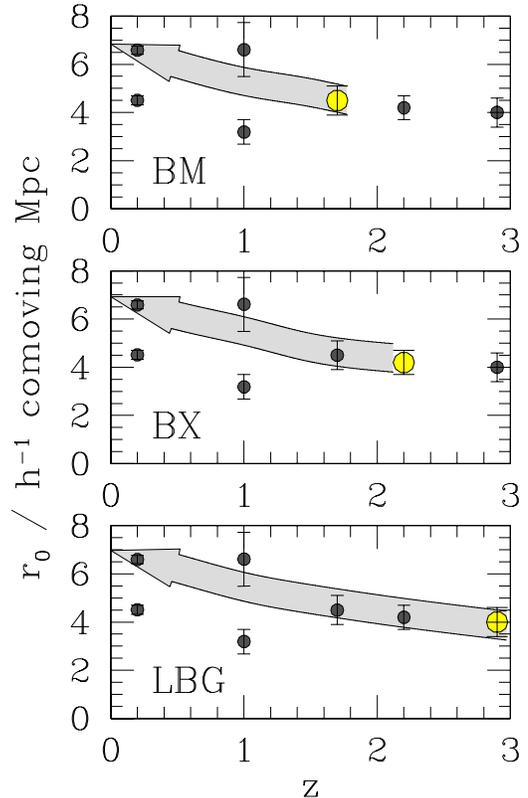}
\caption{
Clustering strength versus redshift.  Top panel:  the large, lightly shaded
point shows the correlation length we infer for galaxies in
the BM sample.  The arrow shows the correlation length these galaxies
will have at lower redshifts. The top edge of the arrow 
shows the evolution of $r_0$ for halos in the LCDM-GIF
simulation that had mass exceeding $M_{\rm top}=10^{12.3} M_\odot$ at the
redshift of identification $z_{\rm id}=1.77$; the bottom
edge is for halos with mass exceeding $M_{\rm bot}=10^{11.9} M_\odot$.
Smaller, darker points show measurements of $r_0$ from other samples:
the SDSS ``elliptical'' and blue samples of Budav\'ari et al. (2003;
$z=0.2$ and $r_0=6.6$, $4.5$ respectively),
the DEEP full and absorption-line samples of Coil et al. (2003;
$z=1$ and $r_0=3.2$, $6.6$ respectively), and the BX and LBG samples
of this paper.  
Middle panel: similar to the top panel, except
$z_{\rm id}=2.32$, $M_{\rm top}=10^{12.2} M_\odot$, $M_{\rm bot}=10^{11.8} M_\odot$, 
and the large point is for the BX sample.
Bottom panel: similar to the top panel, except 
$z_{\rm id}=2.97$, $M_{\rm top}=10^{11.8} M_\odot$, $M_{\rm bot}=10^{11.2} M_\odot$, 
and the large point is for the LBG sample.
\label{fig:r0_vs_z}
}
\end{figure}

The spatial distribution of a population of galaxies
evolves in an easily predictable way as it responds
to the gravitational pull of dark matter.  We used a
simple approach to estimate this evolution from the
LCDM-GIF simulation.
After connecting the observed galaxies to halos
with a range of masses (equation~\ref{eq:mass_scales}),
we measured the evolution in the clustering of those
halos in the simulation and assumed that the galaxies'
clustering would evolve in the same way.
See~\S~\ref{sec:simulated_data}.  Figure~\ref{fig:r0_vs_z}
shows the implied change in correlation length $r_0$ with time 
for the galaxies in our samples.  By $z\sim 2$ galaxies in the
LBG sample will have a correlation length similar to
measured correlation lengths of galaxies in the BM and BX samples.
By $z\sim 1$ their correlation length will be similar to the
correlation length of early-type (i.e., ``absorption line'') galaxies
in the sample of Coil et al. (2003).  By $z\sim 0.2$ their
correlation length will be equal to the observed correlation
length of ellipticals (Budav\'ari et al. 2003).  The evolution
of $r_0$ for galaxies in the BM and BX samples is similar.

\begin{figure}
\plotone{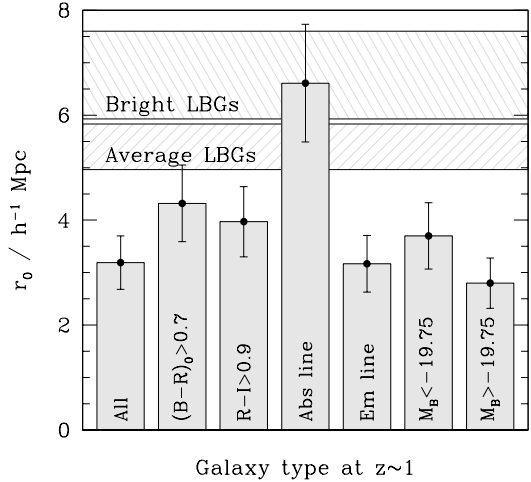}
\caption{
Clustering strength of LBG descendants
and of different galaxy populations in
the $z\sim 1$ DEEP optically selected sample of Coil et al. (2003).
Galaxies at $z\sim 1$ that are brighter ($M_B<-19.75$),
that are redder in observed $R-I$ or rest-frame $(B-R)_0$,
or that have early-type (absorption line) spectra
cluster more strongly than star-forming (emission line) galaxies
or than galaxies in the full sample (``All'').
The shaded regions in the background show the range of
correlation length at $z=1$ expected for descendants of
typical ($23.5<{\cal R}<25.5$) and bright ($23.5<{\cal R}<24.75$) LBGs.  
The correlation lengths
are similar for descendants of BX and BM galaxies (see figure~\ref{fig:r0_vs_z}).
The galaxies at $z\sim 1$ that cluster strongly
enough to be descended from BM, BX, or LBG galaxies
have early type spectra and are red.
\label{fig:z1gals}
}
\end{figure}

\begin{figure}
\plotone{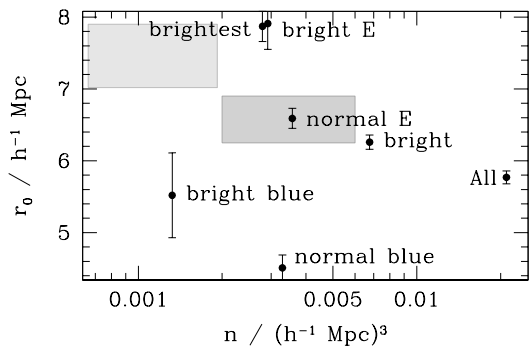}
\caption{Correlation length and number density of 
LBG descendants and of
different
galaxy populations in the SDSS sample of Budav\'ari et al. (2003).
The darker shaded region in the background shows the range of
clustering strength at $z=0.2$ expected for the descendants
of typical LBGs ($23.5<{\cal R}<25.5$)
and the observed number density of typical
LBGs at $z\sim 3$.  The lighter shaded region above and to the left
is for bright LBGs ($23.5<{\cal R}<24.75$).  Values for
BX and BM descendants are similar (see figure~\ref{fig:r0_vs_z}
and equations~\ref{eq:nlbg} and~\ref{eq:nbxbm}).
Points show measured values for galaxy populations at $z\sim 0.2$.
The descriptive labels on the points
correspond to the sub-samples defined by Budav\'ari et al. (2003)
as follows:  
All $\to$ full sample;
bright $\to$ $-21>M_r>-22$; 
brightest $\to$ $-22>M_r>-23$;
normal blue $\to$ $M_r>-21$, $t>0.65$; 
bright blue $\to$ $-21>M_r>-23$, $t>0.65$; 
normal E $\to$ $M_r>-21$, $t<0.02$; 
bright E $\to$ $-21>M_r>-23$, $t<0.02$.
\label{fig:z0gals}
}
\end{figure}

Figures~\ref{fig:z1gals} and~\ref{fig:z0gals} present a more detailed
view of the possible relationships between the descendants
of galaxies in our samples
and various galaxy populations at lower redshift.
Figure~\ref{fig:z1gals} shows that at $z=1$ the descendants'
clustering strength will significantly exceed that of average galaxies
in optical magnitude-limited surveys.  Since these surveys are dominated
by star-forming (``emission-line'') galaxies, we can conclude
that the typical descendant is no longer forming stars by $z\sim 1$.\footnote{
In principle the descendants could have emission-line spectra
if they made up a small part of an otherwise weakly clustered population,
but this possibility seems to be ruled out by the fact that
the number densities of BM/BX/LBG galaxies are similar to the
number density of galaxies in the $z\sim 1$ emission-line sample. (Our estimate 
of the number density in the emission-line sample, shown
in figure~\ref{fig:nhalo}, is from A. Coil, private communication.)}
A similar point was made by Adelberger (2000) and Coil et al. (2004).
The stronger clustering of redder and brighter sub-populations
at $z\sim 1$ is more compatible with the descendants' expected
clustering, but the match is best for the sub-population
with early-type spectra.  This is especially true for
descendants of the brightest galaxies in the BX and LBG samples.
Although the observed number density of early-type galaxies at
redshift $z\sim 1$, roughly $(7\pm 1.5)\times 10^{-3} h^3 {\rm Mpc}^{-3}$
(Chen et al. 2003), is consistent with the idea that most had
a BM/BX/LBG galaxy as a progenitor, we cannot rule out the idea
that some had multiple merged BM/BX/LBG progenitors and others
had none.


Figure~\ref{fig:z0gals} is an analogous plot for redshift $z\sim 0.2$.
Correlation lengths for various populations in
the Sloan Digital Sky Survey (SDSS) were taken
from Budav\'ari et al. (2003).  Number densities were calculated
assuming a surveyed volume of $10^8 h^{-3}$ Mpc$^3$ with
an uncertainty of $\sim 15$\% (T. Budav\'ari 2004, private communication).
The expected $z=0.2$ clustering strength of typical LBG descendants
(darker shaded box) agrees best with the clustering of galaxies
with early-type SEDs
in the Budav\'ari et al. (2003) sample.  The galaxies in our high-redshift 
samples are roughly as numerous as these early-type galaxies,
though the
possibility of merging prevents us from estimating the
number density of their BM/BX/LBG descendants at $z\sim 0.2$.  
The descendants of
brighter LBGs will have a correlation length closer to
that of bright ellipticals, though there are probably not
enough descendants to account for the entire bright elliptical population.

\section{SUMMARY \& DISCUSSION}
\label{sec:summary}
The first part of the paper was concerned with measuring the
spatial clustering of large samples of star-forming galaxies
at redshifts $z\sim 1.7$, $2.2$, and~$2.9$.
We fit a three-dimensional correlation of the
form $\xi(r)=(r/r_0)^{-\gamma}$ to the galaxies' angular clustering
with standard techniques and to the galaxies' redshift clustering
with a new estimator.  The new estimator, $K$, is insensitive to many of the
possible systematic biases in our spectroscopic surveys.
We reached consistent conclusions about the correlation function
with the two approaches, but adopted the angular results since
their random uncertainties were somewhat smaller.
As given in \S~\ref{sec:angularresults},
the best-fit correlation-function parameters from the angular
clustering are
\begin{equation}
r_0/(h^{-1}{\rm Mpc}),\gamma=\cases{ 4.0\pm 0.6, 1.57\pm 0.14& LBG \cr
	                           4.2\pm 0.5, 1.59\pm 0.08& BX  \cr 
	                           4.5\pm 0.6, 1.55\pm 0.10& BM \cr}
\label{eq:correlationlengths}
\end{equation}
where BM, BX, and LBG are the names we have given the $U_nG{\cal R}$
color-selection criteria that we used to find galaxies
at $z\sim 1.7$, $2.2$, and~$2.9$ (\S~\ref{sec:observeddata}).
The quoted $1\sigma$ errors include random uncertainties,
uncertainties in the integral constraint
corrections, and uncertainties in the shapes of the selection functions. 

Since the value of $r_0$
depends on the apparent-magnitude limit of the samples,
at least in the two higher-redshift bins (figure~\ref{fig:clust_seg}),
the reported values of $r_0$ are somewhat arbitrary.
We chose to limit each sample to a fixed range of apparent magnitudes,
$23.5<{\cal R}<25.5$, on the grounds that this resulted in a similar
comoving density in each redshift bin.  Different magnitude limits
would have resulted in different correlation lengths.  Readers
should be aware that the numbers we give are appropriate to the
samples as we have defined them, not to the general galaxy population
at high redshifts.

The second part of the paper was
based on the proposition that
WMAP (Spergel et al. 2003) and other experiments have given
us reliable measurements of the cosmological parameters
and of the shape of the dark matter power-spectrum.
This implies
that we know what sorts of virialized dark-matter halos existed at different
epochs in the past and how their spatial distribution evolved
over time.  Since galaxies reside within dark-matter halos,
they will have the same correlation function as the halos
on large scales (equation~\ref{eq:xigxih}).  The galaxies' clustering
should therefore tell us what sort of halo they reside within.
We found a good match (figures~\ref{fig:xi_halo}
and~\ref{fig:wthet_halo}) between the correlation functions of the galaxies
and of halos with threshold masses ranging from $10^{11.5}M_\odot$ (LBG)
to $10^{12.1}M_\odot$ (BM).\footnote{When we say (for example) that the galaxies
have the same correlation function as halos with threshold mass
$M=10^{11.5}M_\odot$, we mean the subset of
halos with mass $M>10^{11.5}M_\odot$.  The median mass of this
subset is $M=10^{11.86}M_\odot$ in the GIF-LCDM simulation
at $z=2.97$.  Halo subsets
can be defined with schemes more elaborate than our simple 
mass threshold (e.g., Kauffmann et al. 1999,
Bullock, Wechsler, \& Somerville 2002), 
but the differences between the possible schemes are too small to affect
our analysis
when the subsets they produce are constrained to have the same
clustering on large scales.}
Equation~\ref{eq:mass_scales} gives rough $1\sigma$ limits
on the halos' total masses.
Similar masses for Lyman-break galaxies have been
derived with the same approach by
Jing \& Suto (1998), Adelberger et al. (1998), Giavalisco \& Dickinson (2000),
Porciani \& Giavalisco (2002), and others.
Although the estimated masses were derived solely from
the galaxy clustering, they seem reasonable on other grounds.
They cannot be much higher.
The number density of halos would be lower than the number density
of LBGs, for example, if the halo mass were greater than $10^{11.8}M_\odot$.
Such large halo masses would be possible only if significantly more
than one LBG resided in the typical halo, and that is something
that our observations rule out (\S~\ref{sec:correspondence}).
Nor can they be much lower.  The halos that contain LBGs would not have
enough baryons to form the median LBG stellar mass of
roughly $10^{10}M_\odot$
(Shapley et al. 2001\footnote{
This mass is for a Baldry-Glazebrook (2003; eq. 3) IMF, and is
therefore $1.82$ times lower than the value Shapley et al. calculated
for an IMF with a Salpeter slope between 
$0.1$ and $100M_\odot$.  Their assumed Salpeter IMF is probably
unrealistic since the IMF in the solar neighborhood is known
to flatten near $\sim 1M_\odot$ and eventually turn over at lower masses.
See Leitherer (1998) or Renzini (2004)
for further discussion.}) unless their total mass were greater
than about $10^{11}M_\odot$.

We should mention in passing that our best-fit halo masses seem to imply
that only a small fraction of the baryons in the halos are associated
with the observed galaxies.  For example, the best-fit mass threshold
of $10^{11.5}M_\odot$ for LBGs corresponds to a median total mass
of $10^{11.86}M_\odot$ and median baryonic mass of $1.2\times 10^{11}M_\odot$
(for $\Omega_b/\Omega_M\simeq 0.17$, Spergel et al. 2003),
roughly ten times larger than the observed stellar masses of LBGs.
Since the $10^8$ supernovae that explode during the assembly of 
the typical LBG's stellar mass will eject roughly $10^8 M_\odot$ of metals
(e.g., Woosley \& Weaver 1995), enough to enrich at most $1.3\times 10^{10} M_\odot$
of gas to LBGs' typical metallicities of $0.4Z_\odot$ (Pettini et al. 2002), 
their observed interstellar gas cannot contain
a large fraction of the remaining baryons.  These baryons need not be
associated with other objects in the halo, however.  They may
be locked in dim stars that formed in previous episodes 
of star-formation (e.g., Papovich, Dickinson, \& Ferguson 2001),
or may have been heated by various processes to undetectably
high temperatures.  
The latter is presumably the case for nearby galaxies,
whose ratios of mass in stars and gas to total
mass are usually also smaller than the
WMAP value $\Omega_b/\Omega_M\simeq 0.17$.
The Milky Way, for example, has 
a total mass of $10^{12}M_\odot$ (Zaritsky 1999;
Wilkinson \& Evans 1999) and a mass in gas and stars
of only $\sim 8\times 10^{10} M_\odot$ (K. Freeman 2004, private communication),
yet few would assert that its missing baryons belong
to another galaxy in its halo.

After establishing plausible total masses for the halos associated
with the galaxies, we considered some of the implications.
Our arguments were not new (see, e.g., Moustakas \& Somerville 2002,
Martini \& Weinberg 2001, Adelberger et al. 1998).
They seemed worth revisiting only because our knowledge
of the cosmogony, of the local universe, and of high-redshift galaxies
has improved so much in the last few years.  We began by estimating the completeness
of our surveys from a comparison of the galaxies' number densities
to the number densities of halos with similar clustering strength
(figure~\ref{fig:nhalo}).  Similar number densities would imply
that almost all of the most massive halos contained a galaxy that
satisfied our selection criteria; a much lower galaxy number density
would imply that most of the galaxies in massive halos are missed
by our survey.  Defining $\eta$ as the ratio of galaxy to halo
number density, we found rough $1\sigma$ limits of
$0.2<\eta_{\rm LBG}<1$, $0.6<\eta_{\rm BX}<3$,
and $0.5<\eta_{\rm BM}<2.5$.  These limits were derived from the
clustering at radii $r\simgt 1h^{-1}$ comoving Mpc.  The clustering
on smaller scales, sensitive to the possible presence of
more than one galaxy in a halo, implies that the upper limits
on $\eta_{\rm BX}$ and $\eta_{\rm BM}$ should be revised downwards
to $\sim 1.25$.  The data appear consistent
with the claim of Franx et al. (2003) that our selection criteria
find roughly half
of the most massive galaxies at $z\sim 2$.
A completeness of order $50$\% seems plausible to us for other
reasons as well.  Shapley et al. (2001) estimate a lifetime
for the typical LBG of $\sim 3\times 10^8$ yr, for example,
which implies that the typical LBG will be bright enough
for us to detect for only about half of the time that elapsed
between the survey selection limits of $z\sim 3.4$ and $z\sim 2.6$.

We considered next the way the clustering of the galaxies would
evolve (figure~\ref{fig:r0_vs_z}).  Analysis of the GIF-LCDM simulation
suggested that the correlation length of LBG descendants would 
be similar by $z\sim 2.2$ to the correlation length of galaxies
in the BX sample and by $z\sim 1.7$ to the correlation length
of galaxies in the BM sample.  The spatial clustering is therefore
consistent with the idea that we are seeing the same population
at all three redshifts, though the selection criteria's $\sim 50$\%
incompleteness leaves room for the populations to be distinct
and the difference in stellar masses between the LBG ($10^{10}M_\odot$)
and BX ($2\times 10^{10} M_\odot$; Steidel et al. 2005, in preparation)
samples may not be consistent with continuous star-formation at
observed rates through the elapsed time.
Turning our attention to lower redshifts, we found that at $z\sim 1$
our descendants' clustering would most closely match
the observed clustering of galaxies that are red and bright
and have early-type spectra (figure~\ref{fig:z1gals}).   
By $z\sim 0.2$ the estimated clustering of
the descendants suggested elliptical galaxies as the most
likely counterparts (figure~\ref{fig:z0gals}).   The correspondence
is especially hard to dispute for the descendants of the
brightest and most strongly clustered galaxies in the high-redshift samples. 

One conclusion seems difficult to escape:  the descendants of
the galaxies in our samples must have
significantly larger stellar masses than their high-redshift forebears.
Only $\sim 25$\% of the total stellar mass in the local universe
is found in galaxies with stellar masses smaller than
$2\times 10^{10}M_\odot$ (Kauffmann et al. 2003), 
similar to the values in our high-redshift
samples, and these faint galaxies are too weakly clustered to
have descended from the galaxies we find at $1.4<z<3.5$.
Only elliptical galaxies have a spatial distribution consistent
with our expectations for the descendants, and the characteristic
stellar mass of ellipticals is $10^{11}M_\odot$ (Padmanabhan et al. 2004).
The increase in stellar mass from $z\sim 2$ to $z\sim 0$ could have
been produced by ongoing star formation
or by mergers.  Our results at redshift $z\sim 1$ may
favor the latter, but in any case our findings are strongly inconsistent
with traditional notions of monolithic collapse.

It would be unfair to close without mentioning one population
of high-redshift galaxies that we have ignored completely.
These are the bright ($K\simlt 20$) near-infrared-selected galaxies.  
Their reported star-formation rates ($\sim 200M_\odot$/yr),
correlation lengths ($>9 h^{-1}$ Mpc, Daddi et al. 2004)
and stellar masses ($2\times 10^{11}M_\odot$; van Dokkum et al. 2004)
are extraordinary, far larger than the corresponding values
for typical galaxies in our samples.\footnote{These values are
not larger, however, than those for galaxies in our samples
with similar $K$ magnitudes.  See, e.g., Shapley et al. 2004
and Adelberger et al. 2005}
Galaxies with similarly extreme
properties are not a negligible component of the
high redshift universe.
The shape of the $850\mu$m background implies that up to
a third (Cowie, Barger, \& Kneib 2002)
of all stars
could have formed in objects with
star-formation rates greater than $\sim 200 M_\odot$; halos at $z\sim 3$ with 
the large masses $M\simgt 10^{12.7}M_\odot$ implied by $r_0\sim 8$--$9h^{-1}$ Mpc contain
in total almost $20$\% as many baryons as the more numerous
and smaller halos with $M\sim 10^{11.5} M_\odot$ that contain LBGs;
objects with stellar masses $M_\ast>2\times 10^{11}M_\odot$
contain nearly $5$\% of all stars in the local universe
(Kauffmann et al. 2003)
and $20$\% of the stars in local elliptical galaxies (Padmanabhan et al. 2004).
No treatment 
will be entirely complete if it neglects galaxies
similar to those found in near-IR surveys.
The galaxies we studied are neither the most massive, nor the most rapidly
star-forming, nor the most clustered galaxies in the high redshift universe,
but it is precisely this that makes them
plausible progenitors for the early-type galaxies that surround us today.

\bigskip
\bigskip
We are grateful to the Virgo consortium for its public release
of the GIF simulation data and to G. Kauffmann for bringing
the data to our attention.  A. Coil and T. Budav\'ari 
responded helpfully to our questions about their data.
M. Giavalisco, the referee, gave us an insightful report.
KLA, AES, and NAR were supported by
fellowships from the Carnegie Institute of Washington, the Miller Foundation,
and the National Science Foundation.  DKE and CCS were supported
by grant AST 03-07263 from the National Science Foundation.

\end{document}